\documentstyle[12pt,epsfig,amssymb,axodraw]{article}
\begin{document}
\setlength{\parskip}{0.45cm}
\setlength{\baselineskip}{0.75cm}
\begin{titlepage}
\begin{flushright}
DO-TH 98/12 \\ 
DTP/98/36 \\
July 1998 \\
\end{flushright}
\vspace*{-0.3cm}
\begin{center}
\Large
{\bf {Photoproduction of Heavy Quarks}}

\vspace{0.1cm}
{\bf {in Next-to-Leading Order QCD}}

\vspace{0.1cm}
{\bf {with Longitudinally Polarized Initial States}}

\vspace{0.6cm}
\large
I.\ Bojak\\

\vspace*{0.4cm}
\normalsize
{\it Institut f\"{u}r Physik, Universit\"{a}t Dortmund, D-44221 Dortmund,
Germany}

\vspace*{0.1cm}
and\\  

\vspace*{0.3cm}
\large
M.\ Stratmann\\

\vspace*{0.4cm}
\normalsize
{\it Department of Physics, University of Durham, Durham DH1 3LE, England}\\

\large
\vspace*{1.5cm}
{\bf Abstract} \\
\vspace*{-0.3cm}
\end{center}
We present all relevant details of our calculation of the 
complete next-to-leading order $({\cal{O}}(\alpha_s^2 \alpha))$ 
QCD corrections to heavy flavor photoproduction with longitudinally
polarized point-like photons and hadrons. 
In particular we provide analytical results
for the virtual plus soft gluon cross section.
We carefully address the relevance of remaining theoretical uncertainties
by varying, for instance, the factorization and renormalization 
scales independently. Such studies are
of importance for a meaningful first direct determination of 
the polarized gluon density $\Delta g$ from
the total charm production spin asymmetry by the upcoming COMPASS experiment.
It is shown that the scale uncertainty is considerably reduced in 
next-to-leading order, but the dependence on the
charm quark mass is sizable at fixed target energies.
Finally, we study several differential single-inclusive heavy quark
distributions and, for the polarized HERA option,
the total bottom spin asymmetry.
\end{titlepage}
\newpage
%
%
\section{\label{sec:intro}Introduction}

\noindent
Measuring the unpolarized gluon density of the nucleon $g(x,\mu^2)$ at a scale
$\mu$ as a function of the momentum fraction $x$ presents considerable
theoretical and experimental challenges and thus serves as a benchmark for the
steady progress in QCD. 
The determination of $g(x,\mu^2)$ from measurements  of the structure
function $F_2$ in deep inelastic scattering (DIS) is hampered by the
absence of direct couplings to the electroweak probes $(\gamma^*,\,
Z,\, W^{\pm})$.  
However, the increasingly precise $F_2$ data from HERA \cite{ref:hera}
still serve to constrain the small-$x$ behaviour of $g(x,\mu^2)$ 
indirectly in the region $10^{-4}\lesssim x \lesssim 10^{-2}$ with an
accuracy of about $10\%$ \cite{ref:cteq} from the observed scaling violations
$\partial F_2 (x,\mu^2)/\partial \mu^2$.
To determine $g(x,\mu^2)$ over the entire $x$ region, i.e., also at larger
values of $x$, studies of exclusive reactions like direct photon or
di-jet production, where the gluon already enters in leading order (LO),
are indispensable. 
Such measurements are often experimentally much more involved and less 
precise than inclusive DIS. Nevertheless, our knowledge of the unpolarized
gluon density has greatly improved in the past few years (for recent
QCD analyses see \cite{ref:mrst}), but in the region $x\gtrsim 0.1$
the situation is still far from being satisfactory. Here the uncertainty
in $g(x,\mu^2)$ easily amounts to about $100\%$ 
\cite{ref:vv,ref:cteq,ref:mrst}.

Concerning the spin properties of the gluons in a 
longitudinally polarized nucleon, the unpolarized gluon density 
$g(x,\mu^2)$ is defined as the sum of the two possible helicity
distributions, whereas the corresponding {\em polarized} gluon 
density $\Delta g(x,\mu^2)$ is given by the difference. In general we have
for a parton $f$ with $f=g,q,\overline{q}$ 
\begin{eqnarray}
\label{eq:partdef}
{\mathrm unpolarized:} \quad &f(x,\mu^2)& = f_+(x,\mu^2)+f_-(x,\mu^2)\;\;,
\nonumber\\
{\mathrm polarized:} \quad  &\Delta f(x,\mu^2)& =
f_+(x,\mu^2)-f_-(x,\mu^2)\;\;.
\end{eqnarray}
Here $f_+$ and $f_-$ are the densities with the parton spin aligned
and anti-aligned to the spin of the 
nucleon, respectively. In order to measure the two independent
combinations in (\ref{eq:partdef}), we need experimental data and
theoretical calculations distinguishing between different initial
helicity states.  

The long list of spin-dependent DIS experiments \cite{ref:spinexp}
and the recently completed next-to-leading order (NLO)
framework for the evolution of the $\Delta f$ \cite{ref:mertig,ref:werner}
may lead to the expectation that the polarized gluon distribution 
$\Delta g(x,\mu^2)$ should be known with almost similar accuracy 
as $g(x,\mu^2)$ by now.
This is, however, not the case as was revealed by all 
NLO analyses \cite{ref:grsv,ref:gs,ref:pdfs,ref:dss} of presently 
available spin-dependent DIS data. 
In fact it turned out that the $x$-shape of $\Delta g$ is even
almost completely unconstrained. This ignorance is, of course, also
reflected in present values for the first moment of $\Delta g(x,\mu^2)$,
defined by
\begin{equation}
\label{eq:firstmom}
\Delta g(\mu^2)\equiv\int_0^1 dx\, \Delta g(x,\mu^2)\;\;\;,
\end{equation}
which can be estimated at best with an error of $100\%$ for the
time being. $\Delta g(\mu^2)$ plays an important r\^ole in our understanding
of the spin-$1/2$ sum rule for nucleons
\begin{equation}
\label{eq:hsum}
\frac{1}{2}=\frac{1}{2}\Delta\Sigma (\mu^2)+\Delta g(\mu^2)+L_z(\mu^2)\;\;\;,
\end{equation}
where $\Delta \Sigma$ is the total polarization carried by the 
quarks and antiquarks and $L_z$ denotes the sum of the non-perturbative
angular momenta of all partons.

There are three main reasons for the present problems to pin down
$\Delta g(x,\mu^2)$:
\begin{itemize}
\item The measurements of the nucleon spin structure function $g_1$, the 
polarized analogue to the unpolarized structure function $F_1$, 
are still in a ``pre-HERA'' phase. The kinematical coverage of the
fixed target experiments \cite{ref:spinexp}
is by far not sufficient to constrain 
$\Delta g(x,\mu^2)$ from scaling violations 
$\partial g_1(x,\mu^2)/\partial \mu^2$.
\item As already mentioned, the unpolarized gluon density is also constrained
by several exclusive reactions, but corresponding measurements in the
polarized case are still missing.
\item A momentum sum rule for spin-dependent parton densities is lacking, 
i.e., we cannot infer any constraint on $\Delta g$ from the already 
somewhat more precisely known polarized quark distributions. 
In addition, the spin-dependent parton densities $\Delta f$ 
defined in (\ref{eq:partdef}) are not required to be positive definite.
\end{itemize}
Nothing can be done about the last point, of course. 
The small-$x$ region of $g_1$ could be explored at HERA in case that the 
option to longitudinally polarize also the proton beam \cite{ref:herapol} 
will be realized in the future.
First measurements of $\Delta g$ in exclusive reactions will be
provided by the COMPASS fixed target experiment at CERN
\cite{ref:compass} and the BNL RHIC polarized $pp$ collider
\cite{ref:rhic}, which are both currently under construction.

For the determination of the gluon distribution, heavy quark ($Q=c,b$) 
photoproduction 
\begin{equation}
\label{eq:lopgf}
\vec{\gamma} \vec{g} \rightarrow Q \bar{Q}
\end{equation}
is an obvious choice (an arrow denotes a longitudinally polarized
particle from now on). The reconstruction of an open heavy quark state is
experimentally feasible, and in LO only
the photon-gluon fusion (PGF) process in (\ref{eq:lopgf}) contributes, 
which may lead to the hope that an unambiguous  
determination of $\Delta g$ can be performed. 
Thus open charm photoproduction will be used by the 
upcoming COMPASS experiment \cite{ref:compass} to measure $\Delta g$.
All theoretical studies of (\ref{eq:lopgf}) have been performed 
only in LO so far \cite{ref:gr,ref:svhera,ref:others,ref:compass}. 
However, LO estimates usually suffer from a strong dependence on the 
a priori unknown factorization and renormalization scales. Also there
are new NLO subprocesses induced by a light quark replacing the gluon 
in the initial state\footnote{Furthermore, the
on-shell photons in (\ref{eq:lopgf}) cannot only interact directly,
but also via their partonic structure. However, LO estimates of this
unknown ``resolved'' contribution are small for all experimentally
relevant purposes \cite{ref:svhera}.}.
Here the question arises if the PGF
contribution (\ref{eq:lopgf}) still dominates
in the experimentally relevant 
kinematical region as is desirable for a precise determination of $\Delta g$. 
Finally, the NLO corrections have been shown to be sizable
near threshold in the unpolarized case \cite{ref:svn,ref:ellis}.
Clearly, a NLO calculation also for the spin-dependent case is warranted 
in order to provide a meaningful interpretation of the forthcoming 
experimental results. 

This paper provides all relevant details of the 
first calculation of the complete NLO
$({\cal{O}}(\alpha_s^2 \alpha))$ QCD corrections to heavy flavor
photoproduction with point-like photons \cite{ref:letter}. 
In \cite{ref:letter} we only highlighted some
of the most important phenomenological aspects, but we skipped most
calculational details. In addition, we now present, again
for the first time, NLO studies of differential single-inclusive
heavy quark distributions. 
In the next section we will first make some general technical remarks
concerning the polarized calculation. In Section~3 we recall the known LO
results and extend them to $n$ dimensions as is required in course of
the NLO calculation.
In Section~4 we calculate the virtual one-loop corrections to 
(\ref{eq:lopgf}) and examine the gluon bremsstrahlung process 
$\vec{\gamma} \vec{g} \rightarrow Q \bar{Q} g$ in detail. 
Section~5 is devoted to the new genuine NLO contribution with a 
light quark in the initial state 
$\vec{\gamma} \vec{q} \rightarrow Q \bar{Q} q$. The relevant formulae
for calculating total and differential single-inclusive heavy quark
cross sections can be found in Section~6, where we also present
some further phenomenological studies.
Finally, our main results are summarized in
Section~7. In Appendix~A we present the details of the phase space 
calculation. In particular, we focus on 
peculiarities which arise in a polarized calculation using dimensional
regularization. Here we also supply the parametrizations of the 
parton momenta used in our calculations. Appendix~B contains several 
helpful remarks concerning the calculation of the tensor integrals 
needed for the virtual corrections and Appendix~C collects the 
analytical results for the polarized virtual plus soft cross section.

\section{Some General Technical Remarks}
%
In the calculation of the NLO corrections we will encounter the usual
array of ultraviolet (UV), infrared (IR) and mass/collinear (M)
singularities. We choose the framework of $n$-dimensional
regularization to deal with all of these various types of singularities. 
Since our calculations proceed along similar lines as in \cite{ref:svn} for 
the corresponding unpolarized case, we adopt their notation 
$n\equiv 4+\varepsilon$ for the deviation from four space-time
dimensions in order to facilitate comparisons of the intermediate 
results. Of course, all results presented here can be easily converted to
the more common choice $n\equiv4-2\varepsilon$ by just replacing 
$\varepsilon \rightarrow -2\varepsilon$ accordingly.
In the calculation we simply identify the dimension used to
regularize the UV divergencies $n<4$ with the one used for the IR
divergencies $n>4$. Thus we do not distinguish between
$\varepsilon_{UV}<0$ and $\varepsilon_{IR}>0$, which for example leads 
to the following result for the basic loop integral:
\begin{equation}
\label{eq:zero}
\int \frac{d^n q}{q^{2\alpha}}=0\qquad{\mathrm for}\quad\alpha>0\;\;.
\end{equation}
Choosing the $n$-dimensional regularization method introduces some 
complications when polarized processes are investigated due to the 
unavoidable presence of $\gamma_5$ and the totally anti-symmetric 
Levi-Civita tensor $\epsilon_{\mu\nu\rho\sigma}$.
First we shall recall how these quantities appear when projecting onto 
the helicity states of the incoming particles, and then we will 
explain how to deal with them in $n$ dimensions.

One can calculate the squared matrix elements for both unpolarized
and polarized processes {\em simultaneously} using the squared matrix
elements $\left|M\right|^2(h_1,h_2)$ for definite helicities $h_1$ and 
$h_2$ of the incoming particles:
\begin{eqnarray}
\label{eq:unpme}
{\mathrm unpolarized:} \quad \overline{\left| M \right|}^{\: 2} &=&
\frac{1}{2} \left[ \left|M\right|^2(++) + \left|M\right|^2(+-)\right]\;\;,\\
\label{eq:polme}
{\mathrm polarized:} \quad  \Delta\left|M\right|^2 &=&
\frac{1}{2} \left[ \left|M\right|^2(++) - \left|M\right|^2(+-)\right]\;\;.
\end{eqnarray}
This is of course highly desirable, since we obtain an important 
consistency check by comparing with the already known unpolarized results
\cite{ref:svn,ref:ellis}. To obtain $\left|M\right|(h_1,h_2)$
we use the standard helicity projection operators 
(see, e.g., \cite{ref:craigie})
\begin{equation}
\label{eq:polgluon}
\epsilon_{\mu}(k_1,\lambda_1)\, \epsilon^*_{\nu}(k_1,\lambda_1) =
\frac{1}{2} \left[-g_{\mu\nu} + i \lambda_1 \epsilon_{\mu\nu\rho\sigma}
\frac{k_1^{\rho} k_2^{\sigma}}{k_1 \cdot k_2} \right]
\end{equation}
for incoming photons with momentum $k_1$ and helicity $\lambda_1$
(accordingly for gluons with $k_2$ and $\lambda_2$) and
\begin{equation}
\label{eq:polquark}
u(k_2,h) \bar{u}(k_2,h) = \frac{1}{2} \not\! k_2 (1-h \gamma_5)
\end{equation}
for incoming quarks with momentum $k_2$ and helicity $h$
(analogously for antiquarks).

We note that in the {\em{unpolarized}} case one has to average over the $n-2$
spin degrees of freedom for each incoming boson in $n$ dimensions. 
This can be achieved by the replacement 
$-1/2\, g_{\mu\nu} \rightarrow -1/(n-2)\, g_{\mu\nu} = 
- 1/(2+\varepsilon)\, g_{\mu\nu}$
in (\ref{eq:polgluon}) leaving (\ref{eq:unpme}) unchanged.
However, it is convenient, both for the
calculation and for the presentation of the results, to define instead
\begin{equation}
\label{eq:avfac}
E_\varepsilon\equiv\left\{\begin{array}{cr}1/(1+\frac{\varepsilon}{2})&
{\mathrm unpolarized}\\1&{\mathrm polarized}\end{array}\right. .
\end{equation}
One can then perform the unpolarized and polarized calculations
using (\ref{eq:polgluon}), if one multiplies the results
by a factor $E_\varepsilon$ for each incoming boson.
We have also always identified the additional four-vector $\eta^{\sigma}$
usually appearing in (\ref{eq:polgluon}) \cite{ref:craigie} 
to be that of the other incoming particle. 
This is possible since $k_1\cdot k_2 = s/2 \neq 0$ in (\ref{eq:polgluon}) 
and
simplifies the rather lengthy intermediate results considerably.

%
%
\begin{figure}[th]
\vspace*{-0.6cm}
\begin{center}
\setlength{\unitlength}{1pt}
\begin{picture}(330,110)
\Line(0,31.5)(60,31.5)
\Line(0,31.5)(0,36.5)
\Line(60,31.5)(60,26.5)
\ArrowArc(30,31.5)(25,5,90)
\ArrowArc(30,31.5)(25,90,175)
\ArrowArc(30,31.5)(25,185,270)
\ArrowArc(30,31.5)(25,270,355)
\Gluon(30,56.5)(30,41.5){-2.5}{2}
\Gluon(30,21.5)(30,6.5){-2.5}{2}
\GlueArc(30,31.5)(10,10,90){2.5}{2}
\GlueArc(30,31.5)(10,90,170){2.5}{2}
\GlueArc(30,31.5)(10,190,270){2.5}{2}
\GlueArc(30,31.5)(10,270,350){2.5}{2}
\Line(90,31.5)(150,31.5)
\Line(90,31.5)(90,36.5)
\Line(150,31.5)(150,26.5)
\ArrowArc(120,31.5)(25,5,90)
\ArrowArc(120,31.5)(25,90,175)
\ArrowArc(120,31.5)(25,185,270)
\ArrowArc(120,31.5)(25,270,355)
\Gluon(120,56.5)(120,41.5){-2.5}{2}
\Gluon(120,21.5)(120,6.5){-2.5}{2}
\GlueArc(120,31.5)(10,10,90){2.5}{2}
\GlueArc(120,31.5)(10,90,170){2.5}{2}
\GlueArc(120,31.5)(10,190,270){2.5}{2}
\GlueArc(120,31.5)(10,270,350){2.5}{2}
\Line(180,31.5)(240,31.5)
\Line(180,31.5)(180,36.5)
\Line(240,31.5)(240,26.5)
\ArrowArc(210,31.5)(25,5,90)
\ArrowArc(210,31.5)(25,90,175)
\ArrowArc(210,31.5)(25,185,270)
\ArrowArc(210,31.5)(25,270,355)
\Gluon(210,56.5)(210,41.5){-2.5}{2}
\Gluon(210,21.5)(210,6.5){-2.5}{2}
\DashCArc(210,31.5)(10,10,170){3}
\DashCArc(210,31.5)(10,190,350){3}
\ArrowArcn(210,31.5)(15,65,25)
\ArrowArcn(210,31.5)(15,245,205)
\Line(270,31.5)(330,31.5)
\Line(270,31.5)(270,36.5)
\Line(330,31.5)(330,26.5)
\ArrowArc(300,31.5)(25,5,90)
\ArrowArc(300,31.5)(25,90,175)
\ArrowArc(300,31.5)(25,185,270)
\ArrowArc(300,31.5)(25,270,355)
\Gluon(300,56.5)(300,41.5){-2.5}{2}
\Gluon(300,21.5)(300,6.5){-2.5}{2}
\DashCArc(300,31.5)(10,10,170){3}
\DashCArc(300,31.5)(10,190,350){3}
\ArrowArc(300,31.5)(15,25,65)
\ArrowArc(300,31.5)(15,205,245)
\Text(75,31.5)[]{=}
\Text(165,31.5)[]{$-$}
\Text(255,31.5)[]{$-$}
\Text(75,94.5)[]{=}
\Text(165,94.5)[]{+}
\Text(255,94.5)[]{+}
\Text(30,94.5)[]{$P_{\mu\nu}\mapsto\circ$}
\Text(120,94.5)[]{$-g_{\mu\nu}\mapsto\bullet$}
\Text(210,94.5)[]{$\frac{\eta^\mu k^\nu}{\eta\cdot k}$}
\Text(300,94.5)[]{$\frac{k^\mu\eta^\nu}{\eta\cdot k}$}
\Text(15,40)[]{$\circ$}
\Text(45,40)[]{$\circ$}
\Text(105,40)[]{$\bullet$}
\Text(135,40)[]{$\bullet$}
\end{picture}
\end{center}
\caption{\sf \label{fig:ghost} Graphical ``rule'' illustrating 
the replacement of the physical polarization sum 
$P_{\mu\nu}$ $(\circ)$ by $-g_{\mu\nu}$ $(\bullet)$ and appropriate ghost
contributions (dashed lines). The minus signs in the lower half are due to the 
cut ghost loop.}
\end{figure}
As a further simplification one can drop all terms other than 
$-g_{\mu\nu}$ in the symmetric (unpolarized) part of (\ref{eq:polgluon}). 
This of course means that unphysical polarizations 
will be kept in the polarization sums. 
However, unphysical photons decouple completely and unphysical gluons do
not contribute as well, except for those subprocesses where one encounters a
triple-gluon vertex. There one has to introduce incoming external ghost fields
to cancel these unphysical parts \cite{ref:ghost}, when using
$-g_{\mu\nu}$ instead of the physical polarization sum 
$P_{\mu\nu}=\sum_{\lambda}\epsilon_{\mu}(\lambda)\epsilon^*_{\nu}(\lambda)$
\cite{ref:craigie}. Fig.~\ref{fig:ghost}
illustrates this elimination of such terms 
by adding appropriate external ghost contributions.
The extra factors $(-1)$ multiplying each ghost contribution are due to 
the cut ghost loop.

The quantities $\epsilon_{\mu\nu\rho\sigma}$ and $\gamma_5$ introduced 
by (\ref{eq:polgluon}) and (\ref{eq:polquark}), respectively, are of
purely four-dimensional nature and there exists no straightforward
continuation to $n\neq 4$ dimensions. We treat them by applying the
HVBM prescription \cite{ref:hvbm} which provides an internally consistent
extension of $\epsilon_{\mu\nu\rho\sigma}$ and $\gamma_5$ to arbitrary
dimensions. In this scheme the $\epsilon$-tensor continues to be a genuinely
four-dimensional object and $\gamma_5$ is defined as in four
dimensions, implying $\left\{\gamma^\mu,\gamma_5\right\} = 0$ for
$\mu=0,\,1,\,2,\,3$ and $\left[\gamma^\mu,\gamma_5\right]=0$
otherwise. This effectively splits the
$n$-dimensional space into two subspaces, each one equipped with its
own metric: one containing the four
space-time dimensions and one containing the remaining $n-4$ dimensions,
denoted ``hat-space'' henceforth. 
In the matrix elements we then encounter not only conventional
$n$-dimensional scalar products of two momenta, like 
$k\cdot p=g_{\mu\nu} k^\mu p^\nu$, which can be expressed in terms of
the usual Mandelstam variables, but also similar scalar products in the 
hat-space $\widehat{k\cdot p}=\hat{g}_{\mu\nu} \hat{k}^\mu
\hat{p}^\nu$. 

These additional terms would complicate the phase space
calculations considerably, but one can eliminate most of them
by choosing the coordinate system appropriately. 
The three-body phase space integration for the NLO $2\rightarrow 3$
processes exemplifies the problem: Since we are only interested in
single-inclusive heavy quark cross sections in our calculation, it is
possible to assign only the momenta of the two outgoing {\em{unobserved}} 
partons with non-vanishing hat-space contributions. 
The fully measured momenta of the initial states and of the observed
heavy (anti)quark remain purely four-dimensional. 
A convenient choice of coordinates is 
the ``Gottfried-Jackson frame'' \cite{ref:gj,ref:haber}, where the phase space
integration is performed in the rest frame of these two unobserved  
particles. The remaining three momenta can then be oriented in such a way
that they lie in, e.g., the $y-z$ plane. To further facilitate the 
phase space integrations one can finally choose one of these 
three vectors to have only a non-vanishing $z$-component 
(this freedom can be exploited to simplify the angular integrations).  
Using such a choice of coordinates, 
one thus ends up with only {\em one} scalar product 
of hat momenta $\hat{k}^2$, which simplifies the
calculations considerably. In App.~A one can find all required formulae 
concerning the phase space integration in the polarized case 
and the parametrizations of the parton momenta used in our calculations.
For the additional integrals which depend on $\hat{k}^2$ we 
furthermore show in App.~A that thanks to the heavy quark mass they
are all at least of ${\cal O}(\varepsilon)$ and hence drop out 
when the limit $n\rightarrow 4$ is taken in the end.
\begin{samepage}
Thus we arrive at the welcome
conclusion that in the particular case of our calculation hat momenta 
terms in the matrix elements do not contribute to the final result
and can be ignored\footnote{
This is different in calculations involving only {\em{massless}}
particles, as was discussed in \cite{ref:lionel} in the context of 
direct photon production. Notice also that the unphysical helicity
violation at the $qqg$-vertex in the HVBM scheme in $n$ dimensions
(see \cite{ref:werner} for details), is not relevant for our
calculation, since this vertex does not appear
in the mass factorization procedure (see Sections~4 and 5).}.
Concerning all $2\rightarrow 2$ subprocesses presented in Sections~3
to 5 it is then obvious that the same holds
true. Since three of four external particles have their momentum fully 
measured, the fourth is determined by energy-momentum conservation 
and thus all hat components can be eliminated from the calculation.
\end{samepage}
\clearpage

\section{Born Cross Section in $n$ Dimensions}
%
\begin{figure}[ht]
\begin{center}
\setlength{\unitlength}{1pt}
\begin{picture}(342,83)
\put(51,20){
  \begin{picture}(102,63)
    \SetWidth{0.7}
    \Photon(0,56)(51,56){3}{3}
    \Gluon(0,7)(51,7){4}{5}
    \SetWidth{1.8}
    \ArrowLine(102,7)(51,7)
    \ArrowLine(51,7)(51,56)
    \ArrowLine(51,56)(102,56)
  \end{picture}
}
\Text(102,0)[b]{\bf (a)}
\put(189,20) {
  \begin{picture}(102,63)
    \SetWidth{0.7}
    \Photon(0,56)(51,56){3}{3}
    \Gluon(0,7)(51,7){4}{5}
    \SetWidth{1.8}
    \ArrowLine(102,56)(51,56)
    \ArrowLine(51,56)(51,7)
    \ArrowLine(51,7)(102,7)
  \end{picture}
}
\Text(240,0)[b]{\bf (b)}
\end{picture}
\end{center}
\caption{\sf \label{fig:born} Feynman diagrams for the LO photon-gluon fusion
process $\gamma g\rightarrow Q \overline{Q}$.}
\end{figure}
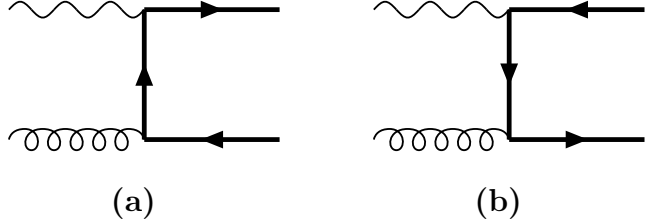
In this section we will recall the well-known LO results for the
unpolarized and polarized photoproduction of heavy flavors. 
Since we encounter $1/\varepsilon^2$ poles in our NLO calculation 
we have to extend these calculations up to ${\cal{O}}(\varepsilon^2)$ 
in $n=4+\varepsilon$ dimensions.
For the contributions to the Born amplitude depicted in Fig.~\ref{fig:born}
we use the following momentum assignment
\begin{equation}
\label{eq:loreac}
\vec{\gamma} (k_1)+\vec{g}(k_2)\rightarrow Q(p_1)+\overline{Q}(p_2)
\end{equation}
and the corresponding Mandelstam variables are given by
\begin{equation}
\label{eq:lomandel}
s=(k_1+k_2)^2,\quad t_1\equiv t-m^2=(k_2-p_2)^2-m^2,
\quad u_1\equiv u-m^2=(k_1-p_2)-m^2\;\;,
\end{equation}
where $s+t_1+u_1=0$, $k_1^2=0$ (``on-shell'' photon),
$k_2^2=0$, and $p_1^2=p_2^2=m^2$ with $m$ denoting the heavy quark mass.
All trace calculations in $n=4+\varepsilon$ dimensions are performed using the
package {\tt TRACER} \cite{ref:tracer}. 
In order to present the unpolarized and polarized results
simultaneously in the most compact form, we will use $|\tilde{M}|^2$ here, 
and in the rest of the paper, to denote {\em{both}} the unpolarized 
$\overline{\left| M \right|}^{\: 2}$ and polarized
$\Delta\left|M\right|^2$ color-averaged squared matrix elements
calculated according to Eqs.~(\ref{eq:unpme})-(\ref{eq:avfac}). 
Similarly, in (\ref{eq:lores}) below
$\tilde{B}_{QED}$ denotes either the unpolarized $B_{QED}$ or
the polarized $\Delta B_{QED}$. 
The LO result can then be expressed as 
\begin{eqnarray}
|\tilde{M}|^2_{\mathrm Born}&=&E_\varepsilon^2\, g_s^2 e^2 e_Q^2
\tilde{B}_{QED}\;\;,\nonumber\\ 
\label{eq:lores}
B_{QED} &=&\frac{t_1}{u_1}+\frac{u_1}{t_1}+
\frac{4 m^2 s}{t_1 u_1}\left(1-\frac{m^2 s}{t_1 u_1}\right)+\varepsilon
\left(\frac{s^2}{t_1 u_1}-1\right)+\varepsilon^2\frac{s^2}{4 t_1 u_1}
\;\;,\\
\Delta B_{QED} &=&\left(\frac{t_1}{u_1}+\frac{u_1}{t_1}\right)
\left(\frac{2 m^2 s}{t_1 u_1}-1\right)\nonumber\;\;,
\end{eqnarray}
where $g_s$ and $e$ are the strong and electromagnetic
coupling constants, respectively, and $e_Q$ is the electromagnetic
charge of the heavy quark in units of $e$, e.g., $e_Q=e_c=2/3$ for
charm quarks.
Notice that the polarized $\Delta B_{QED}$ retains its
four-dimensional form and receives no ${\cal{O}}(\varepsilon)$
contributions in contrast to the unpolarized $B_{QED}$.  

Using the standard $2\rightarrow2$ phase space in $n$ dimensions,
\begin{equation}
\label{eq:ps2}
{\mathrm{dPS}}_2 = \frac{2\pi}{s} \left[(4\pi)^{2+\varepsilon/2}
\Gamma(1+\varepsilon /2)\right]^{-1}
\left(\frac{t_1 u_1-m^2 s}{s}\right)^{\varepsilon /2} dt_1 du_1
\end{equation}
one can then write the $n$-dimensional Born cross section as
\begin{eqnarray}
\label{eq:loxsec}
\frac{d^2\tilde{\sigma}^{(0)}_{g\gamma}}{dt_1du_1}&=&
F_\varepsilon \delta (s+t_1+u_1) |\tilde{M}|^2,\\
F_\varepsilon&\equiv&\frac{\pi}{s^2}\left[(4\pi)^{2+\varepsilon/2}
\Gamma(1+\varepsilon /2)\right]^{-1}
\left(\frac{t_1 u_1-m^2 s}{\mu^2 s}\right)^{\varepsilon /2},\nonumber
\end{eqnarray}
where $F_{\varepsilon}$ collects all phase space factors given in
(\ref{eq:ps2}), the flux factor $1/2s$, and
the mass parameter $\mu$ introduced to keep the gauge couplings 
$g_s$ and $e$ dimensionless in $n$ dimensions.
$\tilde{\sigma}$ denotes the unpolarized and 
polarized cross section  $\sigma$ and $\Delta\sigma$, respectively. 
If one is only interested in the Born result itself, one can of course 
perform the $\varepsilon\rightarrow 0$ limit in (\ref{eq:lores}) 
and simply use $F_{\varepsilon =0}=1/(16 \pi s^2)$. Our four-dimensional
results for $\sigma^{(0)}_{g\gamma}$ and $\Delta\sigma^{(0)}_{g\gamma}$ 
(\ref{eq:loxsec}) agree with those in \cite{ref:lounp,ref:svn} and 
\cite{ref:lopol,ref:conto}, respectively.

\section{\label{sec:gluon}NLO Gluon Contribution}
%
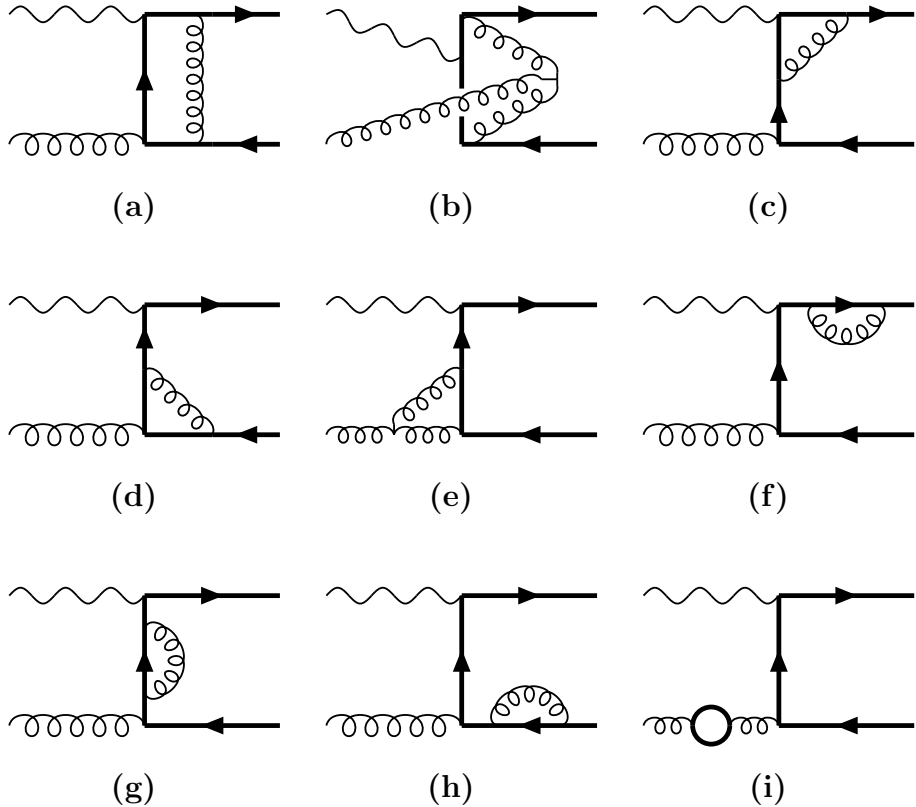
\begin{figure}[th]
\begin{center}
\setlength{\unitlength}{1pt}
\begin{picture}(342,303)
\put(0,240){
  \begin{picture}(102,63)
    \SetWidth{0.7}
    \Photon(0,56)(51,56){3}{3}
    \Gluon(0,7)(51,7){4}{5}
    \Gluon(70,56)(70,7){3}{7}
    \SetWidth{1.8}
    \ArrowLine(102,7)(76.5,7)
    \Line(76.5,7)(51,7)
    \ArrowLine(51,7)(51,56)
    \Line(51,56)(76.5,56)
    \ArrowLine(76.5,56)(102,56)
  \end{picture}
}
\Text(51,220)[b]{\bf (a)}
\put(120,240){
  \begin{picture}(102,63)
    \SetWidth{0.7}
    \Photon(0,56)(51,40){3}{3}
    \Gluon(0,7)(81,31.5){3}{10}
    \Line(81,31.5)(87,31.5)
    \Gluon(51,54)(87,34){3}{4}
    \Line(87,34)(87,31.5)
    \Gluon(87,29)(51,8){3}{4}
    \Line(87,29)(87,31.5)
    \SetWidth{1.8}
    \ArrowLine(102,7)(51,7)
    \Line(51,7)(51,17.5)
    \Line(51,28)(51,56)
    \ArrowLine(51,56)(102,56)
  \end{picture}
}
\Text(171,220)[b]{\bf (b)}
\put(240,240){
  \begin{picture}(102,63)
    \SetWidth{0.7}
    \Photon(0,56)(51,56){3}{3}
    \Gluon(0,7)(51,7){4}{5}
    \Gluon(76.5,56)(51,31.5){3}{4}
    \SetWidth{1.8}
    \ArrowLine(102,7)(51,7)
    \ArrowLine(51,7)(51,31.5)
    \Line(51,31.5)(51,56)
    \Line(51,56)(76.5,56)
    \ArrowLine(76.5,56)(102,56)
  \end{picture}
}
\Text(291,220)[b]{\bf (c)}
\put(0,130){
  \begin{picture}(102,63)
    \SetWidth{0.7}
    \Photon(0,56)(51,56){3}{3}
    \Gluon(0,7)(51,7){4}{5}
    \Gluon(51,31.5)(76.5,7){3}{4}
    \SetWidth{1.8}
    \ArrowLine(102,7)(76.5,7)
    \Line(76.5,7)(51,7)
    \Line(51,7)(51,31.5)
    \ArrowLine(51,31.5)(51,56)
    \ArrowLine(51,56)(102,56)
  \end{picture}
}
\Text(51,110)[b]{\bf (d)}
\put(120,130){
  \begin{picture}(102,63)
    \SetWidth{0.7}
    \Photon(0,56)(51,56){3}{3}
    \Gluon(0,7)(25.5,7){3}{3}
    \Gluon(25.5,7)(51,7){3}{3}
    \Gluon(25.5,11)(51,31.5){3}{4}
    \Line(25.5,7)(25.5,11)
    \SetWidth{1.8}
    \ArrowLine(102,7)(51,7)
    \Line(51,7)(51,31.5)
    \ArrowLine(51,31.5)(51,56)
    \ArrowLine(51,56)(102,56)
  \end{picture}
}
\Text(171,110)[b]{\bf (e)}
\put(240,130){
  \begin{picture}(102,63)
    \SetWidth{0.7}
    \Photon(0,56)(51,56){3}{3}
    \Gluon(0,7)(51,7){4}{5}
    \GlueArc(76.5,56)(12,-180,0){3}{5}
    \SetWidth{1.8}
    \ArrowLine(102,7)(51,7)
    \ArrowLine(51,7)(51,56)
    \ArrowLine(51,56)(102,56)
  \end{picture}
}
\Text(291,110)[b]{\bf (f)}
\put(0,20){
  \begin{picture}(102,63)
    \SetWidth{0.7}
    \Photon(0,56)(51,56){3}{3}
    \Gluon(0,7)(51,7){4}{5}
    \GlueArc(51,31.5)(12,-90,90){3}{5}
    \SetWidth{1.8}
    \ArrowLine(102,7)(51,7)
    \ArrowLine(51,7)(51,56)
    \ArrowLine(51,56)(102,56)
  \end{picture}
}
\Text(51,0)[b]{\bf (g)}
\put(120,20){
  \begin{picture}(102,63)
    \SetWidth{0.7}
    \Photon(0,56)(51,56){3}{3}
    \Gluon(0,7)(51,7){4}{5}
    \GlueArc(76.5,7)(12,0,180){3}{5}
    \SetWidth{1.8}
    \ArrowLine(102,7)(51,7)
    \ArrowLine(51,7)(51,56)
    \ArrowLine(51,56)(102,56)
  \end{picture}
}
\Text(171,0)[b]{\bf (h)}
\put(240,20){
  \begin{picture}(102,63)
    \SetWidth{0.7}
    \Photon(0,56)(51,56){3}{3}
    \Gluon(0,7)(18.5,7){3}{2}
    \Gluon(32.5,7)(51,7){3}{2}
    \SetWidth{1.8}
    \CArc(25.5,7)(7,0,180)
    \CArc(25.5,7)(7,180,360)
    \ArrowLine(102,7)(51,7)
    \ArrowLine(51,7)(51,56)
    \ArrowLine(51,56)(102,56)
  \end{picture}
}
\Text(291,0)[b]{\bf (i)}
\end{picture}
\end{center}
\caption{\sf \label{fig:virt} The NLO virtual corrections to
$\gamma g\rightarrow Q \overline{Q}$.
Reversing the heavy quark lines, except for the non-planar graph (b),
yields the remaining graphs. Massless particle loops similar to graph (i)
vanish, see App.~B.} 
\end{figure}
Next we turn to the NLO corrections to the PGF process
(\ref{eq:lopgf}), where one-loop virtual and gluon bremsstrahlung 
contributions have to be taken into account.
The one-loop virtual corrections displayed in Fig.~\ref{fig:virt} have the
same $2\rightarrow 2$ kinematics as the Born graphs in Sec.~3
and can be also calculated using (\ref{eq:loreac}), (\ref{eq:lomandel}),
(\ref{eq:ps2}), and (\ref{eq:loxsec}). At ${\cal{O}}(\alpha \alpha_s^2)$
only the interference between the virtual $(V)$ and Born $(B)$ amplitudes
of Figs.~\ref{fig:born} and \ref{fig:virt} contributes
\begin{equation}
\label{eq:virtme}
|\tilde{M}|^2_{VB}=2{\mathrm Re} \left(\widetilde{M_V M_B^*}\right)
=E_\varepsilon^2 g_s^4 e^2 e_Q^2 \left[2 C_F \tilde{V}_{QED}+
C_A \tilde{V}_{OK}\right],
\end{equation}
where all quantities with a tilde denote again, as in Eq.~(\ref{eq:lores}),
both the unpolarized and polarized expressions, e.g.,
$\tilde{V}_{QED}$ denotes either $V_{QED}$ or the spin-dependent
$\Delta V_{QED}$.
The color factors associated with the abelian and
non-abelian parts are $C_F=(N_C^2-1)/(2 N_C)$ and $C_A=N_C$,
respectively, where the number of colors is $N_C=3$ for QCD. We note
that $\tilde{V}_{QED}$, which receives contributions
only from the graphs (a), (c), (d) and (f)-(h) in
Fig.~\ref{fig:virt}, corresponds to the process where the gluon is
replaced by a photon in the initial state, i.e., 
$\gamma\gamma\rightarrow Q\bar{Q}$. A complete NLO QCD 
${\cal{O}}(\alpha^2\alpha_s)$ calculation of this process has been
performed recently in \cite{ref:conto} 
for both the unpolarized and polarized case.
Our NLO results for the QED-part of $\gamma g\rightarrow Q\bar{Q}$
agree analytically with the ones presented in \cite{ref:conto}.

In the loop-calculations we encounter Feynman integrals with up 
to four propagators in the denominator. We define the corresponding
one-loop scalar one- ($A_0$), two- ($B_0$), three- ($C_0$) and four-point 
($D_0$) functions as in Ref.~\cite{ref:smith2}, e.g., 
the four-point function needed for the box graphs in Figs.~\ref{fig:virt} 
(a), (b) is defined by
\begin{eqnarray}
\label{eq:fourp}
\lefteqn{D_0(q_1,q_2,q_3,m_1,m_2,m_3,m_4)\equiv}\\
&&\!\!\!\!\mu^{-\varepsilon}
\int\frac{d^n q}{(2\pi)^n}
\frac{1}{(q^2-m_1^2)[(q+q_1)^2-m_2^2][(q+q_1+q_2)^2-m_3^2]
[(q+q_1+q_2+q_3)^2-m_4^2]}\nonumber\;\;,
\end{eqnarray}
where the four external momenta satisfy $q_1+q_2+q_3+q_4=0$ and the $m_i$
are the internal masses. The required scalar integrals are conveniently 
collected in \cite{ref:smith2}, however, 
we have checked them using the standard 
Feynman parametrization techniques. Each fermion propagator and each 
triple-gluon vertex in the loop introduces a loop momentum
$q^{\mu}$ in the numerator. A glance at Fig.~\ref{fig:virt} then reveals 
that the maximal number of loop momenta we face in the numerator is one 
less than the number of propagators, except for graph (i). In particular,
one has to deal with tensorial four-point integrals of first 
($q^{\mu}$) to third ($q^{\mu}q^{\nu}q^{\rho}$) order and with 
tensorial three- and two-point integrals of first ($q^{\mu}$)
and second  ($q^{\mu}q^{\nu}$) order.
We have developed a program which automatically reduces these tensor 
integrals to a set of scalar ones by using an adapted 
Passarino-Veltman decomposition method \cite{ref:pass}, which properly
accounts for all possible $n$-dimensionally regulated divergencies in QCD.
Since this procedure is quite common, we will just mention a few
helpful details in App.~B.

In the virtual cross section UV, IR and M singularities 
show up as $1/\varepsilon$ poles. In the non-abelian $OK$-part
also double poles $1/\varepsilon^2$ occur when IR and M singularities 
coincide. The UV divergencies are removed by the
renormalization procedure, which we implement using the common counterterm
method (``renormalized perturbation theory'').
The counterterms introduce additional contributions similar to those
in Figs.~\ref{fig:virt} (c)-(i), but with the loops replaced by corresponding
renormalization constant dependent ``interactions''. 
For the internal gluon propagators we use the Feynman gauge. As the
renormalization conditions we choose a modified $\overline{\mathrm MS}$
scheme, in which the heavy (anti-)quark is renormalized on-shell and
the light quarks are renormalized using the standard 
$\overline{\mathrm MS}$ prescription. The heavy quark masses are defined
as pole masses. The subtraction for the renormalization of the strong
coupling constant explicitly removes the heavy quark loop contribution 
to the gluon self-energy shown in Fig.~3~(i), see also
(\ref{eq:self}) in App.~B. This leads to a fixed 
flavor scheme with the produced heavy flavor on the one hand and $n_{lf}$
light flavors active in the running of $\alpha_s$ and in the parton
evolution on the other hand \cite{ref:svn,ref:nason2}. The renormalization
constants needed for the construction of the counterterms are then 
calculated to be
\begin{eqnarray}
\label{eq:zets}
Z_m-1&=&\frac{g_s^2 \mu^{-\varepsilon}}{16\pi^2}
C_F 3\left[\frac{2}{\hat{\varepsilon}_m}
-\frac{4}{3}\right]\qquad({\mathrm heavy~quark~mass})\nonumber,\\
Z_2-1&=&-\frac{g_s^2 \mu^{-\varepsilon}}
{16\pi^2} C_F \frac{2}{\hat{\varepsilon}_m}
\qquad({\mathrm heavy~quark~field}),\nonumber\\
Z_3-1&=&\frac{g_s^2 \mu^{-\varepsilon}}
{16\pi^2} \left[(2 C_A-\beta_0)
\frac{2}{\hat{\varepsilon}}+\frac{2}{3}\frac{2}{\hat{\varepsilon}_m}\right]
\qquad({\mathrm gluon~field}),\\
Z_g-1&=&\frac{g_s^2 \mu^{-\varepsilon}}
{32\pi^2} \left[\beta_0
\frac{2}{\hat{\varepsilon}}-\frac{2}{3}\frac{2}{\hat{\varepsilon}_m}\right]
\qquad({\mathrm coupling~constant}),\nonumber\\
Z_{1F}&=&Z_g Z_2 Z_3^{1/2}\qquad({\mathrm quark~gluon~vertex}),\nonumber
\end{eqnarray}
with the QCD beta function $\beta_0\equiv (11 C_A-2 n_{lf})/3$ for the
$n_{lf}$ active light flavors, and we have used the definitions 
$2/\hat{\varepsilon}\equiv 2/\varepsilon+\gamma_E-\ln(4\pi)$ and
$2/\hat{\varepsilon}_m\equiv 2 m^\varepsilon/\hat{\varepsilon}$ with
the Euler constant $\gamma_E$. $Z_{1F}$ is determined to 
${\cal O}(g_s^2)$ using the shown Slavnov-Taylor identity, and the quark
photon vertex renormalization constant can be obtained from $Z_{1F}$ by either
setting $C_A=0$ or by using the QED Ward-Takahashi identity
$Z_{1F}^{QED}=Z_2$ \cite{ref:slav}. The coupling constant $g_s$ in 
(\ref{eq:zets}) and in the matrix element calculations is the renormalized 
one in the counterterm method. We then of course use the renormalization 
group (RG) improved running coupling $g_s(\mu_r^2)$ in the calculation, 
where $\mu_r$ is the renormalization scale at which the subtractions
are performed. 
We have checked that the procedure outlined above is
completely equivalent to the one used in \cite{ref:svn} in the
corresponding unpolarized calculation. In \cite{ref:svn}
the renormalization is performed by replacing the bare mass and strong
coupling constant in the Born cross section with the corresponding 
renormalized quantities (see \cite{ref:svn} for more details). Their
relation between the bare and the renormalized mass can be obtained by
expanding
$m_{\mathrm bare}=Z_m m$ in $\varepsilon$ using the $Z_m$ in 
(\ref{eq:zets}). Analogously, the series in $\varepsilon$ of
$g_s^{\mathrm bare}=Z_g g_s$ using the $Z_g$ of (\ref{eq:zets}) in 
combination with the RG running leads to their relation for the
strong coupling
\begin{equation}
\label{eq:rencoupl}
g_s^{\mathrm bare}\rightarrow g_s(\mu_r^2) 
\left[ 1+ \frac{g_s^2(\mu_r^2)}{32 \pi^2}
\left\{ \left( \frac{2}{\hat{\varepsilon}} + \ln \frac{\mu_r^2}{\mu^2}
\right)\beta_0^f - \frac{2}{3} \ln \frac{m^2}{\mu_r^2} \right\} \right]\;\;,
\end{equation}
where $\beta_0^f$ is defined as $\beta_0$ above, but with
$n_f=n_{lf}+1$ instead of $n_{lf}$ flavors.
The first term in the curly brackets corresponds to the usual
$\overline{\mathrm{MS}}$ prescription and the second one removes the
heavy quark contribution to the gluon self-energy, as already mentioned.

\begin{figure}[th]
\begin{center}
\setlength{\unitlength}{1pt}
\begin{picture}(342,193)
\put(0,130){
  \begin{picture}(102,63)
    \SetWidth{0.7}
    \Photon(0,56)(51,56){3}{3}
    \Gluon(0,7)(51,7){4}{5}
    \Gluon(70,56)(102,31.5){-3}{4}
    \SetWidth{1.8}
    \ArrowLine(102,7)(51,7)
    \ArrowLine(51,7)(51,56)
    \Line(51,56)(76.5,56)
    \ArrowLine(76.5,56)(102,56)
  \end{picture}
}
\Text(51,110)[a]{\bf (a)}
\put(120,130){
  \begin{picture}(102,63)
    \SetWidth{0.7}
    \Photon(0,56)(51,56){3}{3}
    \Gluon(0,7)(51,7){4}{5}
    \Gluon(51,31.5)(102,31.5){3}{6}
    \SetWidth{1.8}
    \ArrowLine(102,7)(51,7)
    \Line(51,7)(51,31.5)
    \ArrowLine(51,31.5)(51,56)
    \ArrowLine(51,56)(102,56)
  \end{picture}
}
\Text(171,110)[b]{\bf (b)}
\put(240,130){
  \begin{picture}(102,63)
    \SetWidth{0.7}
    \Photon(0,56)(51,56){3}{3}
    \Gluon(0,7)(51,7){4}{5}
    \Gluon(70,7)(102,31.5){3}{4}
    \SetWidth{1.8}
    \ArrowLine(102,7)(76.5,7)
    \Line(76.5,7)(51,7)
    \ArrowLine(51,7)(51,56)
    \ArrowLine(51,56)(102,56)
  \end{picture}
}
\Text(291,110)[b]{\bf (c)}
\put(0,20){
  \begin{picture}(102,63)
    \SetWidth{0.7}
    \Photon(0,56)(51,56){3}{3}
    \Gluon(0,7)(51,7){3}{7}
    \Gluon(51,7)(102,7){3}{7}
    \Line(51,7)(51,12)
    \Gluon(51,12)(51,31.5){3}{3}
    \SetWidth{1.8}
    \ArrowLine(102,31.5)(51,31.5)
    \ArrowLine(51,31.5)(51,56)
    \ArrowLine(51,56)(102,56)
  \end{picture}
}
\Text(51,0)[b]{\bf (d)}
\put(120,20){
  \begin{picture}(102,63)
    \SetWidth{0.7}
    \Photon(0,56)(51,56){3}{3}
    \DashArrowLine(0,7)(51,7){5}
    \DashArrowLine(51,7)(102,7){5}
    \Gluon(51,7)(51,31.5){3}{3}
    \SetWidth{1.8}
    \ArrowLine(102,31.5)(51,31.5)
    \ArrowLine(51,31.5)(51,56)
    \ArrowLine(51,56)(102,56)
  \end{picture}
}
\Text(171,0)[b]{\bf (e)}
\put(240,20){
  \begin{picture}(102,63)
    \SetWidth{0.7}
    \Photon(0,56)(51,56){3}{3}
    \DashArrowLine(51,7)(0,7){5}
    \DashArrowLine(102,7)(51,7){5}
    \Gluon(51,7)(51,31.5){3}{3}
    \SetWidth{1.8}
    \ArrowLine(102,31.5)(51,31.5)
    \ArrowLine(51,31.5)(51,56)
    \ArrowLine(51,56)(102,56)
  \end{picture}
}
\Text(291,0)[b]{\bf (f)}
\end{picture}
\end{center}
\caption{\sf \label{fig:brems} Feynman diagrams for the NLO gluon
bremsstrahlung process $\gamma g\rightarrow Q \overline{Q} g$. 
Reversing the heavy quark lines yields the remaining graphs.
In the unpolarized calculation the ghost contributions (e) and (f) 
have to be subtracted to cancel unphysical polarization contributions, 
see the discussion concerning Fig.~\ref{fig:ghost}.}
\end{figure}
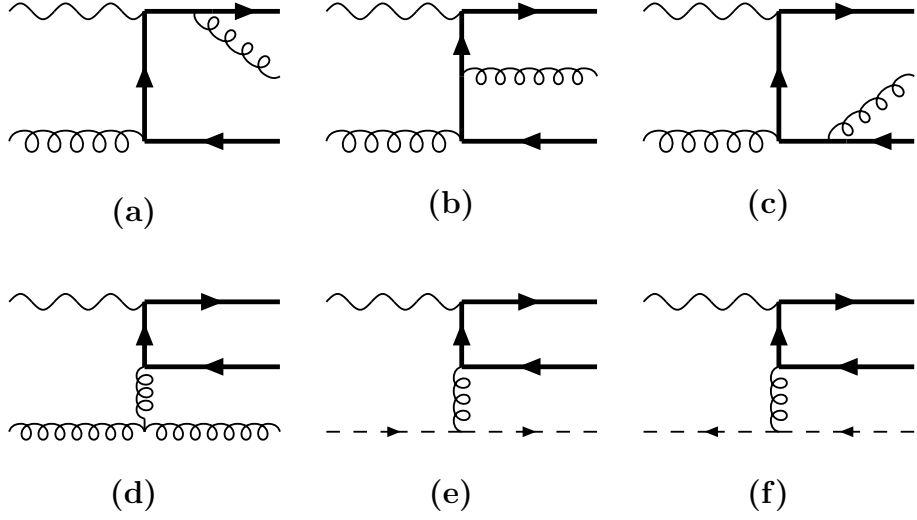
The IR singularities of the virtual cross section are canceled by the 
soft part of the gluon bremsstrahlung contributions. The corresponding 
Feynman diagrams are shown in Fig.~\ref{fig:brems} and the momenta 
assignment is
\begin{equation}
\label{eq:bremsmom}
\vec{\gamma} (k_1)+\vec{g}(k_2)\rightarrow Q(p_1)+\overline{Q}(p_2)
+g(k_3)\;\;.
\end{equation}
In the calculation of this process we keep the kinematical invariants 
as defined in (\ref{eq:lomandel}) and introduce seven additional ones 
\cite{ref:svn}:
\begin{equation}
\label{eq:newmandel}
\renewcommand{\arraystretch}{1.4}
\begin{array}{lll}
s_3=(k_3+p_2)^2-m^2\;\;,\;&s_4=(k_3+p_1)^2-m^2\;\;,\;&s_5=(p_1+p_2)^2=-u_5\;\;,\;\\
t'=(k_2-k_3)^2\;\;,\;&u'=(k_1-k_3)^2\;\;,\;&\\
u_6=(k_2-p_1)^2-m^2\;\;,\;&u_7=(k_1-p_1)^2-m^2\;\;.&
\end{array}
\end{equation}
Of course, only five of these invariants are independent for a $2\rightarrow 3$
process and hence there are many useful relations among the quantities in
(\ref{eq:lomandel}) and (\ref{eq:newmandel}), like
\begin{equation}
\label{eq:manrel}
\renewcommand{\arraystretch}{1.4}
\begin{array}{lll}
u'=-s-u_1-u_7\;\;,\;&t'=-s-t_1-u_6\;\;,\;&u_5=t_1+u_1+s_3\;\;,\;\\
s_3=s+u_6+u_7\;\;,\;&s_4=s+t_1+u_1\;\;.&
\end{array}
\end{equation}

The real gluon bremsstrahlung cross section $(R)$ can then be written as 
\begin{eqnarray}
\label{eq:bremsxsec}
\left(\frac{d^2\tilde{\sigma}^{(1)}_{g\gamma}}{dt_1du_1}\right)^R&=&
F_\varepsilon G_\varepsilon \int d\Omega_\varepsilon |\tilde{M}_R|^2\;\;,\\
G_\varepsilon&\equiv&\frac{\mu^{-\varepsilon}}
{2\pi(4\pi)^{2+\varepsilon/2}} \frac{\Gamma(1+\varepsilon /2)}
{\Gamma (1+\varepsilon)}\frac{s_4^{1+\varepsilon}}
{(s_4+m^2)^{1+\varepsilon/2}}\;\;,\nonumber\\
\int d\Omega_\varepsilon&\equiv&\int_0^\pi d\theta_1
\sin^{1+\varepsilon}\theta_1\int_0^\pi \sin^\varepsilon\theta_2\;\;,
\nonumber
\end{eqnarray}
where we have used the standard $2\rightarrow 3$ phase space (see
App.~A). 
For simplicity we have already replaced the phase space integration over the 
$(n-4)$-dimensional hat-space by ``1'' in (\ref{eq:bremsxsec}), 
since the hat parts do not give any contribution to the final result as 
discussed in Sec.~2 and App.~A. Hence all hat terms in the squared matrix
element $|\tilde{M}_R|^2$ below have to be dropped before 
$|\tilde{M}_R|^2$ is inserted into (\ref{eq:bremsxsec}). 
$G_{\varepsilon}$ parametrizes the differences between the 
$2\rightarrow 2$ prefactor $F_{\varepsilon}$ defined in (\ref{eq:loxsec})
and the $2\rightarrow 3$ prefactor, which is convenient for further 
considerations. 
The squared matrix element can again be
split into an abelian and a non-abelian part
\begin{equation}
\label{eq:bremsme}
|\tilde{M}_R|^2=  \widetilde{M_R M_R^*}
=E_\varepsilon^2 g_s^4 e^2 e_Q^2 \left[2 C_F 
\tilde{R}_{QED}+C_A \tilde{R}_{OK}\right]\;\;.
\end{equation}

Concerning the angular integration $\int d\Omega_\varepsilon$
in (\ref{eq:bremsxsec}) we note
that the Mandelstam variables of (\ref{eq:lomandel}) and
(\ref{eq:newmandel}) are of three distinct types: 
Using the ``set I'' parametrization of the parton momenta in terms of
the angles $\theta_1$ and $\theta_2$ in App.~A, the
Mandelstam variables $t'$ and $u_6$ are of the $[ab]$-type
$a+b\cos\theta_1$, whereas $s_3$, $s_5$, $u'$ and $u_7$ are of the $[ABC]$-type
$A+B\cos\theta_1+C\sin\theta_1\cos\theta_2$. Here $a$, $b$, $A$, $B$, and 
$C$ are functions of the angular-independent ($[\cdot]$-type)
Mandelstam variables $s$, $t_1$, $u_1$, $s_4$
and of the heavy quark mass $m$.
Extensive partial fractioning using the relations given in
(\ref{eq:manrel}) reduces all phase space integrals to the standard form
\begin{equation}
\label{eq:angint}
I_\varepsilon^{(k,l)}=\int d\Omega_\varepsilon (a+b\cos\theta_1)^{-k}
(A+B\cos\theta_1+C\sin\theta_1\cos\theta_2)^{-l}\;\;,
\end{equation}
which can be performed analytically. This reduction procedure can be
completely automatized using a general set of transformation rules
based on relations like in (\ref{eq:manrel}), that is
$[ABC]=[\cdot]+[ab]+[ABC]$. As an example we
demonstrate here two reduction steps for $1/(s_3 u_6 u_7)$
using $s_3=s+u_6+u_7$:
\begin{equation}
\label{eq:redu}
\frac{1}{s_3 u_6 u_7}=\frac{1}{s}\left(\frac{1}{u_6 u_7}
-\frac{1}{u_6 s_3}-\frac{1}{s_3 u_7}\right)\qquad{\mathrm{and}}
\qquad
\frac{1}{s_3 u_7}=\frac{1}{u_6' u_7}-\frac{1}{u_6' s_3}\;\;,
\end{equation}
with $u_6'\equiv s+u_6$. In this way $1/(s_3 u_6 u_7)$ can be
completely reduced to the $[ab][ABC]$ form required in
(\ref{eq:angint}), but the auxiliary $[ab]$
variable $u_6'$ has to be introduced in our example. This always happens
for two $[ABC]$ variables in the denominator.
Though the integrals with this kind of auxiliary variables are as
straightforward as the others, they can also be easily avoided
by introducing a different parametrization
for the parton momenta (``set II'' in App.~A). In this particular
set, $u'$ and $u_7$ are of the $[ab]$-type instead of  $t'$ and $u_6$
for ``set I''. So in our example above, $1/(s_3 u_7)$ would be already of the 
$[ab][ABC]$ type using ``set II'' and no further reduction
would be necessary. 
As a check for the correctness of our phase space calculations, 
we have proceeded in both ways.
It should be remarked that a third
conceivable parametrization of the momenta is not required in our
calculations.

The needed integrals $I^{(k,l)}$ are conveniently
collected in \cite{ref:smith2}. We
have recalculated them and also the few additional ones that occur in
the reduction method using solely ``set I''. The integrals are
straightforward to calculate by using two ``tricks'': Firstly, most
integrals can be derived from a basic one by partial differentiation
with respect to the parameters $a$ and $A$ in (\ref{eq:angint}).
Note that one has to be careful in case of ``collinear'' parameters
$a^2=b^2$ or $A^2=B^2+C^2$, though. Secondly, the transformation 
\begin{equation}
\label{eq:itrans}
I_\varepsilon^{(k,l)}=I_\varepsilon^{(l,k)}(a\leftrightarrow A,
b\rightarrow -\sqrt{B^2+C^2},B\rightarrow \frac{-b B}{\sqrt{B^2+C^2}},
C\rightarrow \frac{-b C}{\sqrt{B^2+C^2}})
\end{equation}
is often helpful. In particular one
can transform ``$A^2=B^2+C^2$'' collinearities into ``$a^2=b^2$'' ones in
this way \cite{ref:wim}. One can also prove
$I_\varepsilon^{(k,l)}(u_6,u_7)=
I_\varepsilon^{(l,k)}(u_6,u_7,t_1\leftrightarrow u_1)$ using this
transformation. 

In order to isolate the divergencies appearing in the soft
$s_4\rightarrow 0$ limit, which cancel the IR singularities 
of the virtual cross section, we examine the
bremsstrahlung result in two regions: that for hard ($s_4>\Delta$) and 
that for soft ($s_4<\Delta$) gluons \cite{ref:svn}. 
Here the auxiliary quantity
$\Delta$ is chosen small enough to be negligible in comparison to $s$, 
$t_1$, $u_1$ and $m^2$. In the hard ($H$) region $\Delta$ 
effectively cuts off the IR singularities, so that only the 
M singularities remain. Thus no double 
poles will be encountered and one needs $G_\varepsilon$ in
(\ref{eq:bremsxsec}) only to ${\cal O}(\varepsilon)$:
\begin{equation}
\label{eq:hardfak}
G_\varepsilon^H\equiv\frac{2(4 \pi)^4 (s_4+m^2)}{s_4}
G_\varepsilon\simeq 16\pi\left[1+\frac{\varepsilon}{2}\left(\gamma_E-
\ln(4 \pi)+\ln\frac{s_4^2}{\mu^2(s_4+m^2)}\right)
\right]\;\;,
\end{equation}
where we have also absorbed some additional factors into
$G_\varepsilon^H$ for convenience. 
The hard gluon cross section then becomes
\begin{eqnarray}
\label{eq:hgxsec}
\left(\frac{d^2\tilde{\sigma}^{(1)}_{g\gamma,OK}}{dt_1du_1}\right)^H&=&
C_A F_\varepsilon G_\varepsilon^H 
E_\varepsilon^2 \alpha_s^2
\alpha e_Q^2 \frac{2}{\varepsilon} 
\tilde{H}_{OK,{\mathrm pole}}+{\cal O}(1)\;\;,\nonumber\\
H_{OK,{\mathrm pole}} &=& \frac{(\rho-1+1/\rho)^2}{s_4}
\left[\frac{t_1}{u_1} \frac{1}{\rho}+\frac{u_1}{t_1} \rho+
\frac{4 m^2 s}{t_1u_1}\left(1-\frac{m^2 s}{t_1u_1}\right)\right]\;\;,\\
\Delta H_{OK,{\mathrm pole}} &=& \frac{2\rho-3+2/\rho}{s_4}
\left(\frac{t_1}{u_1}\frac{1}{\rho}+\frac{u_1}{t_1}\rho\right)
\left(\frac{2 m^2 s}{t_1u_1}-1\right)\;\;,\nonumber
\end{eqnarray}
where only the collinear pole contribution of the non-abelian
$OK$ part is shown.
The hard abelian $QED$ part is completely finite. 
The parameter $\rho\equiv 1-s_4/u_1$
becomes one in the soft limit $s_4\rightarrow 0$, and one can clearly
see the approach to an IR singularity proportional to the Born
$\tilde{B}_{QED}$ of (\ref{eq:lores}).
The finite contributions are too long to be presented
here in an analytical form, but they can be found in our computer program, 
which is available upon request. 
Our unpolarized results agree with those of \cite{ref:svn}.

Turning now to the soft gluon region, we find that one can write the
Mandelstam variables in (\ref{eq:newmandel}) in the
soft limit $s_4\rightarrow 0$ as
\begin{equation}
\label{eq:ms4lim}
\renewcommand{\arraystretch}{1.4}
\begin{array}{lll}
s_3=s_4 \underline{s_3}\;\;,\; &t'=s_4 \underline{t'}\;\;,\;&
 u'=s_4 \underline{u'}\;\;,\\
u_5=-s+s_4 \underline{u_5}\;\;,\; &u_6=u_1+s_4 \underline{u_6}\;\;,\;
& u_7=t_1+s_4 \underline{u_7}\;\;,
\end{array}
\end{equation}
where the underlined quantities are finite dimensionless functions of the
$2\rightarrow 2$ Mandelstam variables (\ref{eq:lomandel}) and $m^2$.
In terms of these variables one can easily
single out the IR singularities by collecting different powers in $s_4$.
Applying the $s_4\rightarrow 0$ limit on the
factor $G_\varepsilon$ in (\ref{eq:bremsxsec}), one obtains a
$s_4^{1+\varepsilon}$ dependence. Thus one only has to keep track of those
parts of the squared matrix element exhibiting a $1/s_4^2$ pole
in the soft limit, since all other terms vanish for
$s_4\rightarrow 0$. In this way one can
easily derive the soft limit of $\tilde{R}_{QED}$ and 
$\tilde{R}_{OK}$ in (\ref{eq:bremsme})
\begin{eqnarray}
\label{eq:softme}
\tilde{S}_{QED}&=&-\frac{2}{s_4^2}\left[ m^2 \left( 1+ 
\frac{1}{\underline{s_3^2}}\right)+
\frac{(2 m^2-s)}{\underline{s_3}}\right] \tilde{B}_{QED},\nonumber\\
\tilde{S}_{OK}&=&\frac{2}{s_4^2}\left[\frac{1}{\underline{t'}}
\left(\frac{t_1}{\underline{s_3}} 
+u_1\right)+\frac{(2 m^2-s)}{\underline{s_3}}\right] \tilde{B}_{QED},
\end{eqnarray}
using the polarized and unpolarized $B_{QED}$ and $\Delta B_{QED}$ in
(\ref{eq:lores}), respectively, in agreement with \cite{ref:svn}.

We rewrite the IR-divergent $s_4^{-1+\varepsilon}$-dependence
in terms of the $\Delta$-distribution\footnote{If $v\equiv 1+ t_1/s$
and $w\equiv -u_1/s+t_1$ are introduced, then 
$s_4\rightarrow 0$ poles show up for $w\rightarrow 1$, 
i.e., $1/s4 \rightarrow 1/(1-w)$. The singular $w\rightarrow 1$
behaviour can then be treated with the usual $+$-distribution
$1/(1-w)_+$ \cite{ref:haber}.}:
\begin{equation}
\label{eq:delplus}
\int_0^\Delta ds_4\, f(s_4) [g(s_4)]_\Delta \equiv \int_0^\Delta ds_4\, 
\left[f(s_4)-f(0)\right] g(s_4)\;\;,
\end{equation}
where $[g(s_4)]_\Delta$ is singular and $f(s_4)$ is finite for
$s_4\rightarrow 0$. 
In particular we need the following identity
\begin{equation}
\label{eq:delident}
s_4^{-1+\varepsilon} = \frac{\Delta^\varepsilon}{\varepsilon}\delta(s_4)+
[s_4^{-1+\varepsilon}]_\Delta\;\;.
\end{equation} 
This yields for the $s_4$-integration of a function ${\cal H}(s_4)$
with a soft pole $s_4^{-1+\varepsilon}{\cal S}(s_4)$ and a finite
${\cal F}(s_4)$ part:
\begin{eqnarray}
\label{eq:s4int}
\lefteqn{\int_0^{s_4^{\mathrm max}} ds_4{\cal H}(s_4)\equiv
\int_0^{s_4^{\mathrm max}} ds_4 \left[s_4^{-1+\varepsilon}
{\cal S}(s_4)+{\cal F}(s_4)\right]}\nonumber\\
&=&\int_0^\Delta ds_4 \left[
\frac{\Delta^\varepsilon}{\varepsilon}\delta (s_4) {\cal S}(s_4)+
\{{\cal S}(s_4)-{\cal S}(0)\} s_4^{-1+\varepsilon}+
{\cal F}(s_4)\right]+
\int_\Delta^{s_4^{\mathrm max}} ds_4 {\cal H}(s_4) \nonumber\\
&\simeq&
\int_0^{s_4^{\mathrm max}} ds_4
\left[\frac{\Delta^\varepsilon}{\varepsilon}\delta (s_4) {\cal S}(s_4)+
\Theta (s_4-\Delta) {\cal H}(s_4)\right]
\end{eqnarray}
where $\Theta$ is the Heavyside step function. In the last step
we have explicitly used that $\Delta$ is small enough to be negligible
with respect to  the $2\rightarrow 2$ Mandelstam variables
(\ref{eq:lomandel}) and $m^2$, taking ${\cal S}(\Delta)\simeq
{\cal S}(0)\neq 0$ and ${\cal F}(\Delta)\simeq {\cal F}(0)=0$ as the
finite limits.

We can thus write the total gluon bremsstrahlung result by multiplying
the hard cross section (\ref{eq:hgxsec}) with $\Theta (s_4-\Delta)$
and adding the soft cross section obtained from (\ref{eq:softme}) with
the $s_4^{-1+\varepsilon}$ replaced by
$\delta (s_4)\Delta^\varepsilon /\varepsilon$ according to 
(\ref{eq:s4int}).
Using this replacement and performing the angular integrations
$d\Omega_\varepsilon$ the soft $(S)$ gluon cross section is then given by
\begin{eqnarray}
\label{eq:sgxsec}
\left(\frac{d^2\tilde{\sigma}^{(1)}_{g\gamma,OK}}{dt_1du_1}\right)^S&=&
C_A F_\varepsilon G_\varepsilon^S E_\varepsilon^2 \alpha_s^2\alpha
e_Q^2\frac{\tilde{B}_{QED}}{2}
\Bigg[\frac{4}{\varepsilon^2}+\frac{2}{\varepsilon}\ln
\frac{t_1}{u_1}+\ln \varkappa\ln\frac{u_1}{t_1}+\frac{1}{2}\ln^2\frac{u_1}{t_1}
+{}\nonumber\\
&&{}-\frac{1}{2}\ln^2 \varkappa
+{\mathrm Li}_2\left(1-\frac{t_1}{\varkappa u_1}\right)-{\mathrm Li}_2
\left(1-\frac{u_1}{\varkappa t_1}\right)
+\frac{2 m^2-s}{s \beta}S(\varkappa)\Bigg]\;\;,\nonumber\\
\left(\frac{d^2\tilde{\sigma}^{(1)}_{g\gamma,QED}}{dt_1du_1}\right)^S&=&
2 C_F F_\varepsilon G_\varepsilon^S E_\varepsilon^2 \alpha_s^2\alpha
e_Q^2\frac{\tilde{B}_{QED}}{2}
\left[-\frac{2}{\varepsilon}+1+\frac{2 m^2-s}{s\beta}
(1-S(\varkappa))\right],
\end{eqnarray}
with
\begin{eqnarray}
\label{eq:unwich}
S(\varkappa)&\equiv&-\frac{2}{\varepsilon}\ln \varkappa+
{\mathrm Li}_2\left(\varkappa^2\right)-
\ln^2 \varkappa+2 \ln \varkappa \ln (1-\varkappa^2)-\zeta (2)\;\;,\nonumber\\
G_\varepsilon^S &\equiv& \frac{2(4 \pi)^4 m^2 \varepsilon}{s_4^2}
G_\varepsilon \rightarrow  16\pi e^{\varepsilon (\gamma_E-\ln (4\pi))/2}
\left(1-\frac{3}{8}\zeta (2)\varepsilon^2\right)
\left(\frac{\Delta^2}{\mu^2 m^2}
\right)^{\varepsilon /2} \delta (s_4),
\end{eqnarray}
and where we have used $\beta\equiv\sqrt{1-4 m^2/s}$, 
$\varkappa\equiv (1-\beta)/(1+\beta)$, the dilogarithm function
${\mathrm Li}_2$ as defined in \cite{ref:duke},
and the Riemann zeta function $\zeta(2)=\pi^2/6$. 
Adding the soft cross section (\ref{eq:sgxsec}) to the renormalized 
virtual cross section obtained from (\ref{eq:virtme}) and
(\ref{eq:zets}) removes 
all IR singularities in the latter, including the $1/\varepsilon^2$ poles.
An additional $1/\varepsilon$ pole in the $OK$ part of 
(\ref{eq:sgxsec}) will be eventually canceled upon adding the soft 
$\delta(1-x)$ contribution of the mass factorization cross section
(\ref{eq:glufact}) discussed below. Our unpolarized results are again 
identical to those of \cite{ref:svn}. In addition we have checked 
that the abelian $QED$ part of the polarized (and unpolarized) result 
is in complete analytical agreement with the NLO
expressions for $\gamma\gamma\rightarrow Q \bar{Q}$ presented in 
\cite{ref:conto}.

To obtain the final result for the gluon cross section,
the remaining M singularities in the hard gluon cross section have to
be removed as well. This can be
achieved by the standard mass factorization procedure. To 
${\cal O}(\alpha_s^2 \alpha)$ 
the reduced finite gluon cross section is given by \cite{ref:svn}
\begin{eqnarray}
\label{eq:glufact}
\lefteqn{\frac{d^2\tilde{\sigma}^{(1)}_{g\gamma}}{dt_1du_1}(\mu_f^2)=
\frac{d^2\tilde{\sigma}^{(1)}_{g\gamma}}{dt_1du_1}(\mu^2)-{}}\\
&&{}-\frac{\alpha_s}{2\pi}\int_0^1 dx \left[
\tilde{P}_{gg}(x)\frac{2}{\varepsilon}+\tilde{F}_{gg}
(x,\mu_f^2,\mu^2)\right] x 
\left[\frac{d^2\tilde{\sigma}^{(0)}_{g\gamma}}{dt_1du_1}\right]
{s\rightarrow x s\choose t_1\rightarrow x t_1}\nonumber\;\;, 
\end{eqnarray}
where $\mu_f$ denotes the factorization scale at which the subtraction
is performed, and $\tilde{F}$ represents the usual freedom in choosing a 
factorization prescription. In the $\overline{\mathrm MS}$ scheme,
which we use, $\tilde{F}$ is given by
\begin{equation}
\label{eq:fmsbar}
\tilde{F}_{ij}(x,\mu_f^2,\mu^2)=\tilde{P}_{ij}(x) \left(\gamma_E-
\ln (4\pi )+\ln\frac{\mu_f^2}{\mu^2}\right)\;\;.
\end{equation}
The $\tilde{P}_{ij}$ in (\ref{eq:glufact}) and (\ref{eq:fmsbar}) are the usual
unpolarized ($P_{ij}$) and polarized ($\Delta P_{ij}$) 
LO Altarelli-Parisi splitting functions \cite{ref:ap} 
\begin{eqnarray}
\label{eq:ggsplit}
P_{gg}(x)&=&\Theta (1-x-\delta ) 2 C_A\left(\frac{1}{1-x}+\frac{1}{x}-2
+x(1-x)\right)+P_{gg}^{\delta}(x)\;\;,\nonumber\\
\Delta P_{gg}(x)&=&\Theta (1-x-\delta ) 2 C_A\left(\frac{1}{1-x}
-2x+1\right)+P_{gg}^{\delta}(x)\;\;,\\
P_{gg}^{\delta}(x)&=&\delta (1-x)\left(\frac{\beta_0}{2}+
2 C_A\ln\delta\right)\;\;.\nonumber
\end{eqnarray}
Since we have regularized all soft singularities in our calculation by
a small parameter $\Delta$ as outlined above, we have to stick to the 
same framework here to deal with the soft $x\rightarrow 1$ divergency
of $\tilde{P}_{gg}$ consistently and cannot simply use the usual
$+$--prescription
$1/(1-x)_+$. In (\ref{eq:ggsplit}) we have thus introduced another
small auxiliary quantity $\delta$ \cite{ref:svn,ref:deltaalt}. 
Of course, $\Delta$
introduced above and $\delta$ are not independent, since they are related
via the mass factorization by $\delta = \Delta/(s+t_1)$. 
$\beta_0$ in (\ref{eq:ggsplit}) includes only the $n_{lf}$ light flavors as 
in (\ref{eq:zets}).

Inserting (\ref{eq:ggsplit}) in (\ref{eq:glufact}) one gets
schematically
\begin{eqnarray}
\label{eq:wich}
\lefteqn{\int_0^1 dx [\delta (1-x) A+\Theta (1-x-\delta) B(x)]
x \delta(x(s+t_1)+u_1) C(xs,xt_1,u_1)}\\
&=&\delta(s_4) A C(s,t_1,u_1)+\Theta (s_4-\Delta) \left[-u_1
B\left(\frac{-u_1}{s+t_1}\right) C\left(\frac{-u_1 s}{s+t_1},
\frac{-u_1 t_1}{s+t_1},u_1\right)\right]\;\;,\nonumber
\end{eqnarray}
where we have explicitly used the relation between $\delta$ 
and $\Delta$. Thus the contribution from 
mass factorization naturally splits into a soft and a hard part, which
can be added to the corresponding cross sections.
As already mentioned, the $2/\varepsilon$
pole in the soft $\delta(1-x)$
part of (\ref{eq:glufact}) cancels the remaining
pole in the $OK$ part of the soft cross section (\ref{eq:sgxsec}).
In the hard cross section (\ref{eq:hgxsec})
the $2/\varepsilon$ collinear pole is removed and one is left
over with the finite contributions from the pole part 
obtained from the $\varepsilon$-expansion of the prefactor. 
Examining this factor (\ref{eq:hardfak}) and our choice for the
$\overline{\mathrm{MS}}$ factorization scheme in (\ref{eq:fmsbar}), 
it is easy to see that the reduced hard gluon cross section can be simply 
obtained from (\ref{eq:hgxsec}) by setting 
\begin{equation}
\label{eq:redhard}
F_\varepsilon \tilde{G}_\varepsilon^H \rightarrow 
\frac{1}{s^2}\;\;,\qquad E_\varepsilon\rightarrow 1\;\;,
\qquad{\mathrm and}\qquad
\frac{2}{\varepsilon}\rightarrow \ln\frac{s_4^2}{m^2(s_4+m^2)}
-\ln\frac{\mu_f^2}{m^2}\;\;, 
\end{equation}
where one can now perform the $\varepsilon\rightarrow 0$ limit.
Again we agree with the unpolarized reduced hard gluon cross section 
of \cite{ref:svn}.

To complete the calculation of the gluon cross section, we now add the 
$\delta(s_4)$ mass factorization contribution in (\ref{eq:glufact}), see
Eq.~(\ref{eq:wich}), to the renormalized virtual plus
soft part $(V+S)$. We write the result in three parts using the usual abelian
and non-abelian split and, in addition, separating off the part proportional to
$\beta_0/2$. The latter piece vanishes if one identifies the
renormalization scale
with the factorization scale, i.e., $\mu_r=\mu_f$, as is usually done:
\begin{equation}
\label{eq:vssplit}
\left(\frac{d^2\hat{\sigma}^{(1)}_{g\gamma}}{dt_1du_1}\right)^{\mathrm V+S}
=\frac{\alpha_s^2 \alpha e_Q^2}{s^2} 
\left[2 C_F \left(\tilde{L}_{QED}+\tilde{L}_{QED}^\Delta\right)+
C_A \left(\tilde{L}_{OK}+\tilde{L}_{OK}^\Delta\right)+
\frac{\beta_0}{2} \tilde{L}_{RF}\right]\;\;.
\end{equation}
The $\tilde{L}^\Delta$ explicitly depend on the auxiliary quantity
$\Delta$. The polarized $\Delta L$ are presented in App.~C and the
unpolarized $L$ are in complete agreement with those obtainable
from App.~A of \cite{ref:svn} and App.~D of \cite{ref:smith2}. 
The numerical treatment of the $\tilde{L}^\Delta$ terms is 
discussed in App.~C. 

To conclude this section we note that the presented results have been
calculated for a detected heavy {\em antiquark} in the final state, because
the heavy quark was integrated out in the calculations. But since all
gluon matrix elements are symmetric with respect to
$p_1\leftrightarrow p_2$, the same gluon cross section holds for a
detected heavy quark as well. On the other hand there is an
{\em asymmetry} in the
non-abelian part of the gluon cross section with respect to 
$k_1 \leftrightarrow k_2$, since the outgoing gluon with momentum
$k_3$ can only ``couple'' to the incoming gluon with momentum $k_2$, but
not to the photon with momentum $k_1$.

\section{\label{sec:quark}NLO Light Quark Contribution}
%
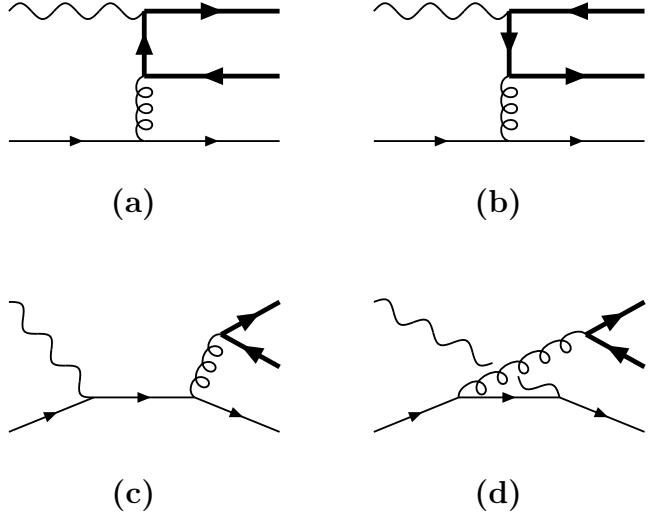
\begin{figure}[th]
\vspace*{-0.6cm}
\begin{center}
\setlength{\unitlength}{1pt}
\begin{picture}(342,193)
\put(51,130){
  \begin{picture}(102,63)
    \SetWidth{0.7}
    \Photon(0,56)(51,56){3}{3}
    \ArrowLine(0,7)(51,7)
    \ArrowLine(51,7)(102,7)
    \Gluon(51,7)(51,31.5){3}{3}
    \SetWidth{1.8}
    \ArrowLine(102,31.5)(51,31.5)
    \ArrowLine(51,31.5)(51,56)
    \ArrowLine(51,56)(102,56)
  \end{picture}
}
\Text(102,110)[b]{\bf (a)}
\put(189,130){
  \begin{picture}(102,63)
    \SetWidth{0.7}
    \Photon(0,56)(51,56){3}{3}
    \ArrowLine(0,7)(51,7)
    \ArrowLine(51,7)(102,7)
    \Gluon(51,7)(51,31.5){3}{3}
    \SetWidth{1.8}
    \ArrowLine(51,31.5)(102,31.5)
    \ArrowLine(51,56)(51,31.5)
    \ArrowLine(102,56)(51,56)
  \end{picture}
}
\Text(240,110)[b]{\bf (b)}
\put(51,20){
  \begin{picture}(102,63)
    \SetWidth{0.7}
    \Photon(0,56)(32,20){3}{3}
    \Gluon(70,20)(80,43.75){3}{3}
    \ArrowLine(0,7)(32,20)
    \ArrowLine(32,20)(70,20)
    \ArrowLine(70,20)(102,7)
    \SetWidth{1.8}
    \ArrowLine(102,31.5)(80,43.75)
    \ArrowLine(80,43.75)(102,56)
  \end{picture}
}
\Text(102,0)[b]{\bf (c)}
\put(189,20){
  \begin{picture}(102,63)
    \SetWidth{0.7}
    \Photon(0,56)(44.7,33){3}{3}
    \Photon(54.4,28)(70,20){-3}{1}
    \Gluon(32,20)(80,43.75){3}{5}
    \ArrowLine(0,7)(32,20)
    \ArrowLine(32,20)(70,20)
    \ArrowLine(70,20)(102,7)
    \SetWidth{1.8}
    \ArrowLine(102,31.5)(80,43.75)
    \ArrowLine(80,43.75)(102,56)
  \end{picture}
}
\Text(240,0)[b]{\bf (d)}
\end{picture}
\end{center}
\caption{\sf \label{fig:qini} Feynman diagrams for the  NLO light quark 
initiated process $\gamma q\rightarrow Q \overline{Q} q$.} 
\end{figure}
In NLO one encounters a new type of subprocess with a light (anti)quark in
the initial state
\begin{equation}
\label{eq:qmom}
\vec{\gamma} (k_1)+\vec{q}(k_2)\rightarrow Q(p_1)+\overline{Q}(p_2)
+q(k_3)\;\;,
\end{equation}
which can be calculated along the same lines as the gluon bremsstrahlung
contribution in the previous section using 
Eqs.~(\ref{eq:newmandel})-(\ref{eq:bremsxsec}). The squared matrix
element calculated from the graphs shown in Fig.~\ref{fig:qini}
can be decomposed according to whether the photon couples to the heavy
quark with charge $e_Q$ (in units of $e$) in the
``Bethe-Heitler-graphs''  (a) and (b), or to the light quark with
charge $e_q$, as for the ``Compton-graphs'' (c) and (d): 
\begin{equation}
\label{eq:qme}
|\tilde{M}_q|^2= \widetilde{M_q M_q^*}
=E_\varepsilon g_s^4 e^2 \frac{C_F}{2}
\left[e_Q^2 \tilde{A}_1+e_q^2 \tilde{A}_2+
e_q e_Q \tilde{A}_3\right]\;\;,
\end{equation}
where $\tilde{A}_3$ denotes the interference contribution of both types 
of processes.
Notice that since we have now only one boson in the
initial state, the photon, only one factor $E_\varepsilon$ appears in
(\ref{eq:qme}). 
Since this production mechanism appears for the first time in NLO and
there are no gluons in the final state, we do not encounter IR singularities 
in the calculation. All single poles can be solely
attributed to collinear configurations, and thus an expansion
to ${\cal O}(\varepsilon )$ is sufficient here as for the
hard gluon cross section in Eqs.~(\ref{eq:hardfak}) and
(\ref{eq:hgxsec}). 

The phase space integrations and the preceding partial fractioning 
proceed just as was explained in Sec.~4 and so we can immediately 
quote the results here
\begin{eqnarray}
\label{eq:qxsec}
\frac{d^2\tilde{\sigma}^{(1)}_{q\gamma,A_i}}{dt_1du_1}&=&
\frac{C_F}{2} F_\varepsilon G_\varepsilon^H 
E_\varepsilon \alpha_s^2 \alpha e_i^2
\frac{2}{\varepsilon} \tilde{A}_{i}^{\mathrm pole}+{\cal O}(1)\;\;,\nonumber\\
A_{1}^{\mathrm pole} &=&-\frac{2(\rho-1)+1/\rho}{u_1}
\left[\frac{t_1}{u_1}\frac{1}{\rho}+\frac{u_1}{t_1}\rho+\frac{4 m^2 s}{t_1u_1}
\left(1-\frac{m^2s}{t_1u_1}\right)\right]\;\;,\nonumber\\
\Delta A_{1}^{\mathrm pole} &=&-\frac{2-1/\rho}{u_1}
\left(\frac{t_1}{u_1}\frac{1}{\rho}+\frac{u_1}{t_1}\rho\right)
\left(\frac{2 m^2 s}{t_1u_1}-1\right)\;\;,\\
A_{2}^{\mathrm pole} &=&-\frac{\tau-2+2/\tau}{t_1}
\left[\frac{\tau^2 t_1^2+u_1^2}{\tau s^2}+\frac{2 m^2}{s}\right]\;\;
,\nonumber\\
\Delta A_{2}^{\mathrm pole} &=&-\frac{2-\tau}{t_1}
\left[-\frac{\tau^2 t_1^2+u_1^2}{\tau s^2}-\frac{2 m^2}{s}\right]\;\;,\nonumber
\end{eqnarray}
where again only the collinear pole contributions 
are given and $e_i^2$ denotes $e_Q^2$, $e_q^2$, and $e_Q e_q$ for
$i=1,\,2,$ and 3, respectively.
The interference contribution $\tilde{A}_3$ is completely finite. 
The new parameter $\tau\equiv 1-s_4/t_1$ becomes one in the limit 
$s_4\rightarrow 0$, exactly like $\rho\equiv 1-s_4/u_1$ already 
introduced in (\ref{eq:hgxsec}).
$\tilde{A}_1^{\mathrm{pole}}$ stems from $1/t'$ terms in the
matrix element (\ref{eq:qme}).
$t'$ becomes zero for the collinear configuration $k_3=k_2 (1-x)$. In
``set I''  this zero shows up as $a=-b$ with $a=\rho u_1 s_4/[2(s_4+m^2)]$.
Note that $\tilde{A}_1^{\mathrm{pole}}$ becomes proportional to
$B_{QED}$ in the limit $k_3\rightarrow 0$, and thus $s_4\rightarrow 0$,
for $x\rightarrow 1$.
Similarly, $\tilde{A}_2^{\mathrm{pole}}$ originates from $1/u'$ terms.
$u'$ becomes zero for $k_3=k_1 (1-x)$ and in ``set I'' one finds
$A=-\sqrt{B^2+C^2}$ with $A=\tau t_1 s_4/[2(s_4+m^2)]$.
In the limit $x\rightarrow 1$,  $\tilde{A}_2^{\mathrm{pole}}$
becomes proportional to $A_{QED}$, which is defined in
(\ref{eq:qqbar}) below. The complete expressions for this
subprocess are too long to be given here but can be found in our
computer program. Our unpolarized results fully agree with those
of \cite{ref:svn}.

Again the mass factorization procedure removes the collinear
singularities. To ${\cal O}(\alpha_s^2 \alpha)$
the reduced finite quark cross section is given by \cite{ref:svn}
\begin{eqnarray}
\label{eq:qfact}
\lefteqn{\frac{d^2\hat{\sigma}^{(1)}_{q\gamma}}{dt_1du_1}(\mu_f^2)=
\frac{d^2\tilde{\sigma}^{(1)}_{q\gamma}}{dt_1du_1}(\mu^2)-{}}\\
&&{}-\frac{\alpha_s}{2\pi}\int_0^1 dx_1 \left[
\tilde{P}_{gq}(x_1)\frac{2}{\varepsilon}+\tilde{F}_{gq}
(x_1,\mu_f^2,\mu^2)\right] x_1 
\left[\frac{d^2\tilde{\sigma}^{(0)}_{g\gamma}}{dt_1du_1}\right]
{s\rightarrow x_1 s\choose t_1\rightarrow x_1 t_1}-{}\nonumber\\ 
&&{}-\frac{\alpha}{2\pi}\int_0^1 dx_2 \left[
\tilde{P}_{q\gamma}(x_2)\frac{2}{\varepsilon}+\tilde{F}_{q\gamma}
(x_2,\mu_f^2,\mu^2)\right] x_2 
\left[\frac{d^2\tilde{\sigma}^{(0)}_{q\overline{q}}}{dt_1du_1}
\right]{s\rightarrow x_2 s\choose u_1\rightarrow x_2 u_1}\nonumber\;\;, 
\end{eqnarray}
and for light antiquarks the analogous relation with
$q\rightarrow\overline{q}$ holds.
The first subtraction in (\ref{eq:qfact}) corresponds to the collinear
configuration in the Bethe-Heitler part, whereas the second piece
refers to the collinear Compton contribution. The quark-gluon 
\begin{equation}
\label{eq:qgsplit}
P_{gq}=C_F \left[\frac{1+(1-x)^2}{x}\right],\qquad
\Delta P_{gq}=C_F (2-x)\;\;,
\end{equation}
and the photon-quark splitting functions
\begin{equation}
\label{eq:pqsplit}
P_{q\gamma}=e_q^2 N_C \left[x^2+(1-x)^2\right]\;\;,\qquad
\Delta P_{q\gamma}=e_q^2 N_C (2 x-1)\;\;,
\end{equation}
in (\ref{eq:qfact}) can be obtained from \cite{ref:ap}. 
The corresponding antiquark splitting functions are identical
and $\tilde{F}_{ij}$ was already specified in (\ref{eq:fmsbar}) in the
$\overline{\mathrm{MS}}$ scheme.

It should be noted that the subtraction term proportional to
$\tilde{P}_{q\gamma}$ in (\ref{eq:qfact}) implicitly introduces 
the quark content of the real (on-shell) photon. 
Clearly, one cannot obtain a factorization scheme independent result
taking into account only the 
``direct'' point-like photon contribution without adding the
corresponding ``resolved'' cross section which probes the parton content of
the photon. This is made evident by the appearance of
$\tilde{F}_{q\gamma}$ in (\ref{eq:qfact}), allowing for arbitrary
redefinitions of the factorization scheme (i.e., of the photonic
quark densities in NLO), which can only be compensated by the NLO resolved 
contributions. Since the spin-dependent resolved
cross section has not been calculated in NLO yet, it has to be estimated in 
LO. A further complication arises here, because the parton content of
longitudinally,
i.e., circularly, polarized photons is experimentally completely
unknown for the time being, and one has to rely on realistic models
\cite{ref:models} when estimating the size of the resolved contribution. 
However, it has been demonstrated in \cite{ref:svhera} that even for
large spin-dependent photonic densities, the ``background'' from
resolved photon reactions should be very small for all experimentally
relevant total or differential cross sections. In particular this
is the case at fixed target energies, as for COMPASS. Only for the
total charm production spin asymmetry at collider energies the resolved
contribution can be as large as $30\%$ \cite{ref:svhera}. But in this
kinematical region, charm production anyway suffers from large
statistical errors and appears to be unmeasurable at the polarized
HERA option \cite{ref:herapol}.

It should also be remarked that for NLO photonic parton densities, 
unpolarized \cite{ref:grvphot} as well as polarized \cite{ref:svphot} ones,
often the so-called $\mathrm{DIS}_\gamma$ factorization scheme 
\cite{ref:grvdis} rather
than the $\overline{\mathrm{MS}}$ prescription is used, since it provides a
better perturbative stability between LO and NLO
quark densities. In this case one
either has to transform the the densities back to the
$\overline{\mathrm{MS}}$ scheme \cite{ref:grvphot,ref:svphot}
before using them in the calculation
of the NLO resolved contribution or one has to use the
appropriate $\mathrm{DIS}_\gamma$ expression for $\tilde{F}_{q\gamma}$ 
in (\ref{eq:qfact}), see the Appendix of \cite{ref:grvdis}.   

To calculate the factorization contribution for the Compton part
proportional to $e_q^2$ in (\ref{eq:qfact}), the 
Born cross section for the $q\overline{q}\rightarrow Q\overline{Q}$ process 
in $n$ dimensions is required.
%
\begin{figure}[th]
\vspace*{-0.6cm}
\begin{center}
\setlength{\unitlength}{1pt}
\begin{picture}(102,49)
  \SetWidth{0.7}
  \ArrowLine(0,0)(34,24.5)
  \ArrowLine(34,24.5)(0,49)
  \Gluon(34,24.5)(68,24.5){4}{4}
  \SetWidth{1.8}
  \ArrowLine(102,0)(68,24.5)
  \ArrowLine(68,24.5)(102,49)
\end{picture}
\end{center}
\caption{\sf \label{fig:qqbar} The LO quark-antiquark annihilation process  
$q \overline q\rightarrow Q \overline{Q}$.}
\end{figure}
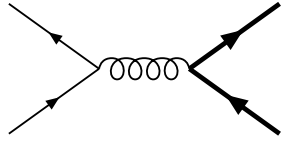
Only one diagram contributes here, see Fig.~\ref{fig:qqbar},
which can be easily calculated using the general
$2\rightarrow 2$ phase space expression of (\ref{eq:loxsec})
and the result reads
\begin{eqnarray}
\label{eq:qqbar}
\frac{d^2\tilde{\sigma}^{(0)}_{q\overline{q}}}{dt_1du_1}&=&
F_\varepsilon \frac{C_F}{N_C} g_s^4 \tilde{A}_{QED},\nonumber\\
A_{QED} &=&\frac{t_1^2+u_1^2}{s^2}
+\frac{2 m^2}{s}+\frac{\varepsilon}{2}\;\;,\\
\Delta A_{QED} &=&-\frac{t_1^2+u_1^2}{s^2}
-\frac{2 m^2}{s}+\frac{\varepsilon}{2}\;\;.\nonumber
\end{eqnarray}
The mass factorization in (\ref{eq:qfact}) is again performed 
with the $\overline{\mathrm MS}$ 
prescription (\ref{eq:fmsbar}), and the finite reduced quark cross
section can be obtained by applying Eq.~(\ref{eq:redhard}) to
(\ref{eq:qxsec}).
Our results fully agree with \cite{ref:svn} in the unpolarized case again.

Finally, it is important to point out that $\tilde{A}_1$ and $\tilde{A}_2$ in
(\ref{eq:qxsec}) stay unchanged for $p_1\leftrightarrow p_2$, whereas
$\tilde{A}_3$ changes sign. Thus if one wants to observe a heavy quark
instead of a  heavy antiquark, one can use 
$[e_Q^2 \tilde{A}_1+e_q^2 \tilde{A}_2-e_q e_Q \tilde{A}_3]$ 
in (\ref{eq:qme}) with the same expressions for the $\tilde{A}_i$.
In addition charge conjugation implies that
\begin{equation}
\label{eq:qcc}
d\hat{\sigma}\left(\overline{q}\gamma\rightarrow\overline{Q}\right)=
d\hat{\sigma}\left(q\gamma\rightarrow Q\right)\qquad{\mathrm and}\qquad
d\hat{\sigma}\left(\overline{q}\gamma\rightarrow Q\right)=
d\hat{\sigma}\left(q\gamma\rightarrow\overline{Q}\right)\;\;.
\end{equation}
Thus one can use the same $\tilde{A}_i$ for the contribution due to a incoming
antiquark in (\ref{eq:qmom}) as well, taking into account a 
negative sign for $\tilde{A}_3$. Note 
that the sign change of $\tilde{A}_3$ also implies that $\tilde{A}_3$ does 
not contribute to the total cross section 
(\ref{eq:partonxsec}) below, since the result cannot depend on
whether the heavy quark or heavy antiquark is integrated first.

\section{Hadronic Cross Sections and Numerical Results}
%
Let us now turn to some phenomenological aspects. The total photon-parton
cross section can be expressed in terms of scaling functions in both the 
unpolarized and polarized case $(i=g,\,q,\,\bar{q})$ \cite{ref:svn,ref:letter}:
\begin{eqnarray}
\label{eq:partonxsec}
\tilde{\hat{\sigma}}_{i\gamma}^Q (s,m^2,\mu_f,\mu_r) &=&
\int_{s(1-\beta)/2}^{s(1+\beta)/2}d(-t_1)\int_{-m^2 s/t_1}^{s+t_1}
d(-u_1) \frac{d^2\tilde{\hat{\sigma}}_{i \gamma}(s,t_1,u_1)}{dt_1du_1}\\
&=&\frac{\alpha\alpha_s}{m^2} \left[ 
\tilde{f}_{i\gamma}^{(0)}(\eta) + 4\pi \alpha_s 
\left\{\tilde{f}_{i\gamma}^{(1)}(\eta) +\;
\tilde{\!\!\bar{f}}_{i\gamma}^{(1)}(\eta) \ln \frac{\mu_f^2}{m^2}\right\}
\right]\nonumber\;\;,
\end{eqnarray}
where $\eta\equiv s/(4 m^2)-1$ and $\beta$ is defined below 
Eq.~(\ref{eq:unwich}). 
$\tilde{f}^{(0)}_{i \gamma}$ and $\tilde{f}^{(1)}_{i \gamma}$,
$\;\tilde{\bar{\!\!f}}^{(1)}_{i \gamma}$ stand for the LO and NLO corrections,
respectively\footnote{Note that we still use the ``tilde notation'' as
a shorthand to denote both the longitudinally polarized and unpolarized
cross sections simultaneously.}. This coefficient functions can be further
decomposed depending on the electric charge of the heavy and light quarks,
$e_Q$ and $e_q$, respectively:
\begin{eqnarray}
\label{eq:fg}
\tilde{f}_{g \gamma} (\eta) &=& e_Q^2 \tilde{c}_{g \gamma} (\eta)\;\;, \\
\label{eq:fq}
\tilde{f}_{q \gamma} (\eta) &=& e_Q^2 \tilde{c}_{q \gamma} (\eta) +
                                e_q^2 \tilde{d}_{q \gamma} (\eta)\;\;,
\end{eqnarray}
with corresponding expressions for the $\;\tilde{\bar{\!\!f}}_{i \gamma}$.
Note that the interference contribution proportional to $e_Q e_q$ drops out
in $\tilde{f}_{q \gamma}$ as discussed in Sec.~5.

The behaviour of the spin-dependent coefficient functions 
(\ref{eq:fg}) and (\ref{eq:fq}) has been
already shown and discussed in detail in \cite{ref:letter} for the
conventional choice $\mu_f=\mu_r$. 
Here we just want to point
out that for $\mu_f\neq \mu_r$ one receives an extra contribution from
$\tilde{L}_{RF}$ in (\ref{eq:vssplit}), see also Eq.~(\ref{eq:lrf}) in 
App.~C.
This can be easily accounted for by adding an
appropriate term to the NLO gluonic coefficient function
\begin{equation}
\label{eq:gkchange}
\tilde{f}_{g\gamma}^{(1)}(\eta)=e_H^2 \tilde{c}_{g\gamma}^{(1)}(\eta)+
\frac{\beta_0}{16\pi^2} \tilde{c}_{g\gamma}^{(0)}(\eta)
\ln\frac{\mu_r^2}{\mu_f^2}\;\;,
\end{equation}
and, of course, by using $\mu_r$ as the scale for $\alpha_s$. Notice that 
this is equivalent to the replacement
\begin{equation}
\label{eq:alpchange}
\alpha_s(\mu_r^2)\rightarrow\alpha_s(\mu_r^2) 
\left(1+\alpha_s(\mu_r^2) \frac{\beta_0}{4\pi}
\ln\frac{\mu_r^2}{\mu_f^2}\right)\;\;, 
\end{equation}
in Eq.~(\ref{eq:partonxsec}), keeping only terms up to 
${\cal O}(\alpha_s^2\alpha)$.
We will study the effect of varying $\mu_f$ and $\mu_r$ independently
on the total hadronic heavy flavor photoproduction cross section
given by
\begin{equation}
\label{eq:totalxsec}
\tilde{\sigma}^Q_{\gamma p}(S,m^2,\mu_f,\mu_r) =
\sum_{f=g,q,\bar{q}} \int_{4m^2/S_{\gamma p}}^1 dx\, \tilde{\hat{\sigma}}
(x S,m^2,\mu_f,\mu_r) \tilde{f}(x,\mu_f^2)
\end{equation}
in detail below.
$S$ in (\ref{eq:totalxsec}) 
denotes the available photon-hadron c.m.s.\ energy and
the $\tilde{f}$ are the (un)polarized parton distributions.

Before that let us first of all recall the relevant formulae for 
calculating differential single-inclusive heavy (anti)quark distributions. We 
denote the momenta in the photon-hadron cross section by 
\begin{equation}
\label{eq:hadphot}
\gamma(k_1)+H(K_2)\rightarrow \overline{Q}(p_2) \left[Q(p_1)\right]+X,
\end{equation}
and use the following hadronic invariants for the observed 
heavy antiquark
\begin{equation}
\label{eq:hadmandel}
S=(k_1+K_2)^2=\frac{s}{x}\;,\;
T_1=(K_2-p_2)^2-m^2=\frac{t_1}{x}\;,\;
U_1=(k_1-p_1)^2-m^2=u_1\;,
\end{equation}
where we have introduced the momentum fraction $x$ in $k_2=x K_2$ 
to relate the hadronic to the partonic
variables in Eq.~(\ref{eq:lomandel}). For an observed heavy quark one
would exchange $p_1\leftrightarrow p_2$ in (\ref{eq:hadmandel})
and there would be a $t_1\leftrightarrow u_1$ crossing. 

The hadronic single-inclusive heavy (anti)quark cross section reads
\begin{equation}
\label{eq:spics}
\frac{d^2\tilde{\sigma}_{H\gamma}^Q}{dT_1dU_1}=
\sum_{g,q,\bar{q}}\int_{x_{\mathrm min}}^1 dx x \tilde{f}(x,\mu_f^2)
\frac{d^2\tilde{\hat{\sigma}}_{f\gamma}}{dt_1du_1}\;\;,
\end{equation}
and the lower limit $x_{\mathrm min}$ of the integration 
is determined from
\begin{equation}
\label{eq:xmin}
s_4=x S+x T_1+U_1\geq \Delta \Rightarrow 
x_{\mathrm min}=\frac{\Delta-U_1}{S+T_1}\;\;.
\end{equation}
For the actual integrations it is convenient to change the variable
from $x$ to $s_4$ in (\ref{eq:spics}) with the limits
$\Delta\leq s_4 \leq s_4^{\mathrm max}= S+T_1+U_1$:
\begin{equation}
\int_{x_{\mathrm{min}}}^1 dx= \frac{1}{S+T_1}
\int_\Delta^{s_4^{\mathrm{max}}} ds_4 \;\;,
\end{equation}
where $\Delta$ is the cutoff introduced above to separate the hard
and the soft gluon cross sections. The soft plus virtual gluon cross section
proportional to $\delta (s_4)$ has to be evaluated with elastic
kinematics $(s+t_1+u_1=0)$. However, for numerical purposes we rewrite
the $\ln^i \Delta/m^2$ $(i=0,\,1,\,2)$ terms in $d\tilde{\sigma}^{S+V}$
into integrations over $s_4$ as outlined in App.~C. In this way
the soft plus virtual and the hard parts of the gluonic cross section
can be directly added. One can always set $\Delta =0$ for the
light-quark induced subprocess.

The differential heavy (anti)quark cross section (\ref{eq:spics}) should
be expressed in variables more suited for experimental measurements:
\begin{eqnarray}
\label{eq:expvar}
{\mathrm transverse~momentum/mass}~ &x_T&\equiv
\frac{p_T}{p_T^{\mathrm{max}}}\;\;,\qquad
m_T^2\equiv m^2+p_T^2=\frac{T_1U_1}{S}\;\;,\nonumber\\
{\mathrm rapidity:}~ &y&\equiv{\mathrm artanh}\frac{p_L}{E}=
\frac{1}{2}\ln\frac{U_1}{T_1}\;\;,\\
{\mathrm Feynman}-x:~ &x_F&\equiv \frac{p_L}{p_L^{\mathrm max}}=
\frac{1}{\beta_S} \frac{T_1-U_1}{S}\nonumber\;\;,
\end{eqnarray}
where $\beta_S\equiv\sqrt{1-4 m^2/S}$. The energy and the
longitudinal momentum of the heavy antiquark are
given by $E=m_T \cosh y$ and $p_L=m_T \sinh y$, respectively.
$p_T=|\vec{p}_T|$ is the absolute size of the transverse
momentum and
$p_T^{\mathrm{max}}=p_L^{\mathrm{max}}=\sqrt{S}\beta_S/2$.
$y$ and $x_F$ of the observed $\bar{Q}$ in (\ref{eq:expvar}) are defined
in the hadronic c.m.s.\ with the forward direction ($y>0$) along
the incoming photon, i.e., 
\begin{equation}
T_1=-\sqrt{S}m_T e^{-y}=-\sqrt{S}p_L^{\mathrm{max}}(\chi-x_F)\;,\;\; 
U_1=-\sqrt{S}m_T e^{y}=-\sqrt{S}p_L^{\mathrm{max}}(\chi+x_F)\;,
\end{equation}
where $\chi\equiv\sqrt{x_F^2+(m_T/p_L^{\mathrm{max}})^2}$.

The Jacobians needed to express
(\ref{eq:spics}) in the variables (\ref{eq:expvar}) are 
\begin{equation}
dT_1 dU_1=S dm_T^2 dy=\frac{S}{\chi}dm_T^2 dx_F\;\;,
\end{equation}
and $dm_T^2=2 x_T (p_T^{\mathrm{max}})^2 dx_T$, etc.
By integrating the variables in (\ref{eq:expvar}) over the appropriate limits
\begin{eqnarray}
\label{eq:igrenz}
S\int_{m^2}^{S/4} dm_T^2 
\int_{-{\mathrm arcosh}\frac{\sqrt{S}}{2 m_T}}^{{\mathrm arcosh}
\frac{\sqrt{S}}{2 m_T}} dy &=&
S\int_{-\frac{1}{2} \ln \frac{1+\beta_S}{1-\beta_S}}^{
\frac{1}{2} \ln \frac{1+\beta_S}{1-\beta_S}} dy
\int_{m^2}^{\frac{S}{4 \cosh^2 y}}dm_T^2\;\;,\nonumber\\
S\int_{m^2}^{S/4} dm_T^2
\int_{-\frac{1}{\beta_S}\sqrt{1-\frac{4 m_T^2}{S}}}^{\frac{1}{\beta_S}
\sqrt{1-\frac{4 m_T^2}{S}}}\frac{dx_F}{\chi}
&=&S\int_{-1}^1 dx_F \int_{m^2}^{\frac{S}{4}(1-\beta_S^2 x_F^2)}
\frac{dm_T^2}{\chi}
\end{eqnarray}
the total cross section (\ref{eq:totalxsec}) is of course recovered.

Finally it should be noted that experiments do not determine 
the (differential) longitudinally polarized cross section 
$(d)\Delta\sigma$ itself, but rather the corresponding spin asymmetry
\begin{equation}
\label{eq:asym}
A_{\gamma H}^Q = \frac{(d) \Delta \sigma}{(d)\sigma}\;\;.
\end{equation}
In (\ref{eq:asym}), which is nothing but the counting rate asymmetry
for the two possible helicity alignments of the incoming photon and
hadron in analogy to Eqs.~(\ref{eq:unpme}) and (\ref{eq:polme}),
the experimental normalization
uncertainty and other systematical errors conveniently drop out.
However, in the following we will concentrate on the polarized cross
section itself, since we are mainly interested in the influence of the
spin-dependent NLO corrections.
The calculation of the spin asymmetry
(\ref{eq:asym}) would introduce additional theoretical
uncertainties associated with the {\em unpolarized}
(differential) cross section. 

Equipped with the necessary technical framework, we now turn to some
numerical applications. Unless otherwise stated we use here the
GRV \cite{ref:grv} and GRSV \cite{ref:grsv} ``standard'' set of unpolarized
and longitudinally polarized parton distributions, respectively,
in our calculations. The strong dependence of the results on the chosen,
experimentally poorly constrained polarized gluon distribution $\Delta g$
(and to a lesser extent also on the unpolarized gluon $g$) has been
already demonstrated in \cite{ref:letter} for the case of the total
charm production cross section and the corresponding spin asymmetry.
Of course this sensitivity in turn implies that such a measurement
is particularly suited to pin down $\Delta g$ more precisely. 
Unfortunately we have no data so far, but in the near future 
COMPASS \cite{ref:compass} is going to
measure the total charm spin asymmetry $A^c_{\gamma p}$ with 
sufficient accuracy \cite{ref:compass,ref:letter}. Therefore we
mainly focus on the
kinematical range accessible by COMPASS in our analyses below, i.e.,
$\sqrt{S}= \sqrt{S_{\gamma p}}= 10\,\mathrm{GeV}$.
It is currently under scrutiny whether it is physically feasible and sensible
to run HERA in a polarized collider mode in the future \cite{ref:herapol},
and therefore we either show or comment on the corresponding
results at HERA collider
energies $\sqrt{S}\simeq 100-300\,\mathrm{GeV}$ as well.

%
%
\begin{figure}[t]
\begin{center}
\vspace{-1.4cm}
\epsfig{file=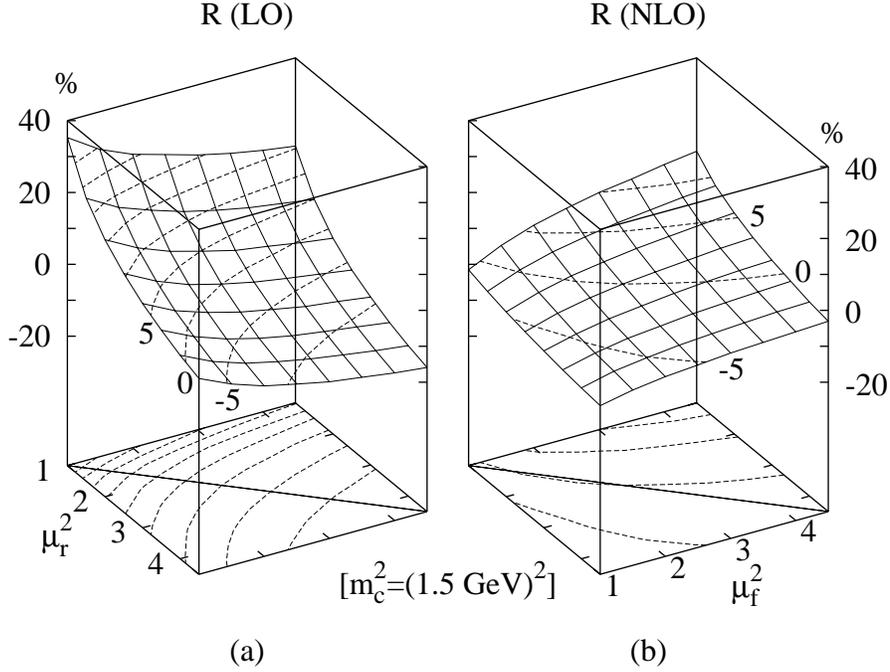}
\end{center}
\vspace{-0.7cm}
\caption{\sf \label{fig:murmuf} Renormalization $\mu_r$
and factorization $\mu_f$ scale dependence of the LO (a) and NLO (b) 
$R$ as defined in Eq.~(\ref{eq:deviation}) in percent 
for $\sqrt{S}=10\;\mathrm{GeV}$.
$\mu_r$ and $\mu_f$ are in units of the charm
quark mass $m_c=1.5\;\mathrm{GeV}$. 
The contour lines are in steps of $5\%$ and for convenience a line
corresponding to the usual choice $\mu_r=\mu_f$ is shown at the base of the
plots.}
\end{figure}
In \cite{ref:letter} we kept the renormalization and factorization scales
equal at the common choice $\mu_f=\mu_r=2m$ for the total charm spin
asymmetry. Here we investigate the 
theoretical uncertainty induced by varying $\mu_f$ and $\mu_r$
independently in the range $\mu_f^2,\,\mu_r^2=m^2,\ldots,4.5m^2$ with
$m=1.5\,\mathrm{GeV}$ for the charm quark mass.
Fig.~7 shows the deviation $R$ of the LO (a) and NLO (b) polarized
total cross sections $\Delta \sigma_{\gamma p}^c$ (\ref{eq:totalxsec}) from the
results obtained for the choice $\mu_f^2=\mu_r^2=2.5m^2$, i.e.,
\begin{equation}
\label{eq:deviation}
R=\frac{\Delta \sigma^c_{\gamma p}(\mu_r^2,\mu_f^2) -
        \Delta \sigma^c_{\gamma p}(\mu_r^2=\mu_f^2=2.5m^2)}
       {\Delta \sigma^c_{\gamma p}(\mu_r^2=\mu_f^2=2.5m^2)}
\end{equation}
for $\sqrt{S}=10\,\mathrm{GeV}$ in percent.
The contour lines are in steps of $5\%$, and at the base of the plot 
the line corresponding to the usual choice $\mu_f=\mu_r$ 
is shown for convenience.
As can be inferred from comparing Figs.~7(a) and (b),
the scale dependence has been drastically reduced in NLO 
in the entire range for $\mu_f$ and $\mu_r$,
which underlines the usefulness of the NLO results. 
Moreover, in NLO the choice $\mu_f=\mu_r$ is approximately on the
contour for $R=0$, and R is flattest for large
$\mu_f$ and $\mu_r$. This a posteriori motivates and justifies our
choice of scales, $\mu_f=\mu_r=2 m$, in \cite{ref:letter}. 
For reasonable changes of $\mu_f$ and $\mu_r$
in Fig.~7, the polarized total charm production
cross section (\ref{eq:totalxsec}) varies by about $\pm 10 \%$ in NLO 
as compared to about $-15 \%$ to $35 \%$ in LO.
It should be noted that one finds very similar results also for a
higher c.m.s.\ energy of, e.g., $\sqrt{S}=100\,\mathrm{GeV}$.

%
\begin{figure}[t]
\begin{center}
\vspace{-0.1cm}
\epsfig{file=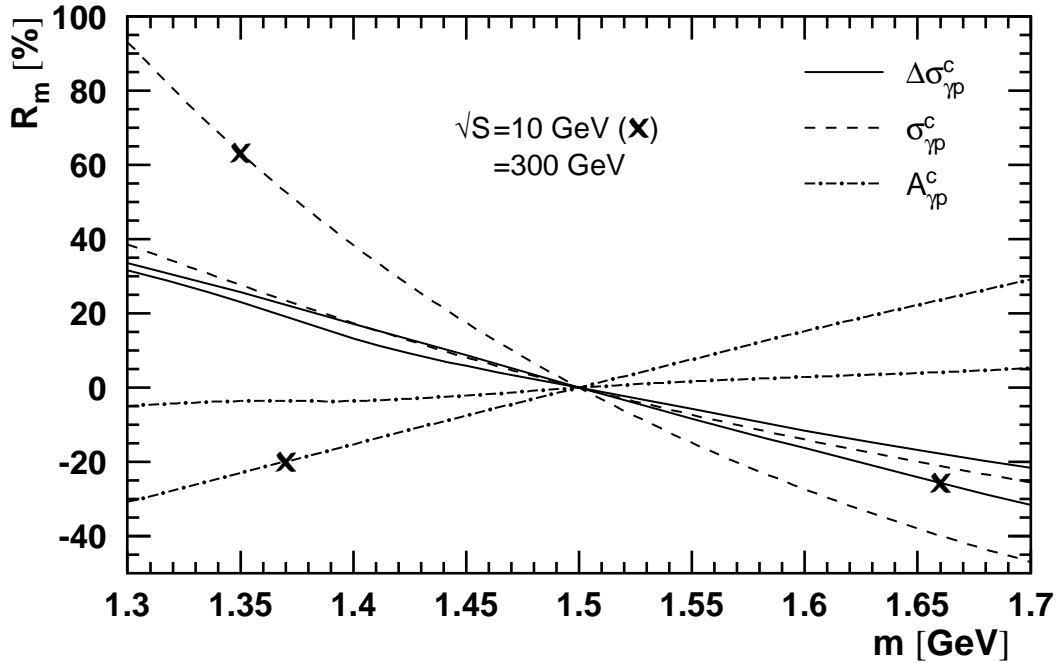}
\end{center}
\vspace{-0.3cm}
\caption{\sf \label{fig:mc} Mass dependence 
of the (un)polarized total charm photoproduction cross 
section (\ref{eq:totalxsec})
and the spin asymmetry (\ref{eq:asym}) in LO and NLO in terms of the
ratio $R_m$ as defined in the text for 
$\sqrt{S}=10\,\mathrm{GeV}$ (denoted by a cross) and 300 GeV.}
\end{figure}
Fig.~8 illustrates in a similar fashion as in Fig.~7 the dependence
of the polarized and unpolarized total charm photoproduction cross sections
(\ref{eq:totalxsec})
and the corresponding longitudinal spin asymmetry (\ref{eq:asym})
on the mass of the
charm quark for two values of $\sqrt{S}$. 
We vary $m$ around our standard choice $m=1.5\;{\mathrm GeV}$ 
by $\pm 0.2\;{\mathrm GeV}$, and $R_m$ in Fig.~8 
is defined in analogy to Eq.~(\ref{eq:deviation}). 
As is expected, far above the production threshold
for $\sqrt{S}=300~{\mathrm{GeV}}\gg 4m^2$,
the dependence of the quantities in Fig.~8 on
the precise value of $m$ is strongly reduced as compared to the 
results obtained for $\sqrt{S}=10\,\mathrm{GeV}$.
For a reliable extraction of $\Delta g$ by
COMPASS, the mass uncertainty is far more important than
the scale dependence in Fig.~7, since the experimentally relevant
spin asymmetry $A^c_{\gamma p}$ varies by as much as $30\%$ in the
shown mass range. A good determination of $m$
is thus mandatory for a meaningful determination of $\Delta g$ at low
energies, not too far above threshold. In addition, we have already
stressed in \cite{ref:letter}, that for a determination of $\Delta g$ at 
fixed target energies further complications arise also from our poor 
knowlegde of the unpolarized gluon distribution for $x\gtrsim 0.1$.
Unfortunately, a measurement of $A_{\gamma p}^c$ at collider
energies, where the theoretical uncertainties are much better under
control than at low energies, appears to be not feasible, 
since $A_{\gamma p}^c$ is at best of the same size as the expected
statistical errors for such a measurement \cite{ref:svhera}. This also does
not improve for $p_T$ or $y$ differential charm distributions. 

%
%
\begin{figure}[t]
\begin{center}
\vspace{-2.3cm}
\epsfig{file=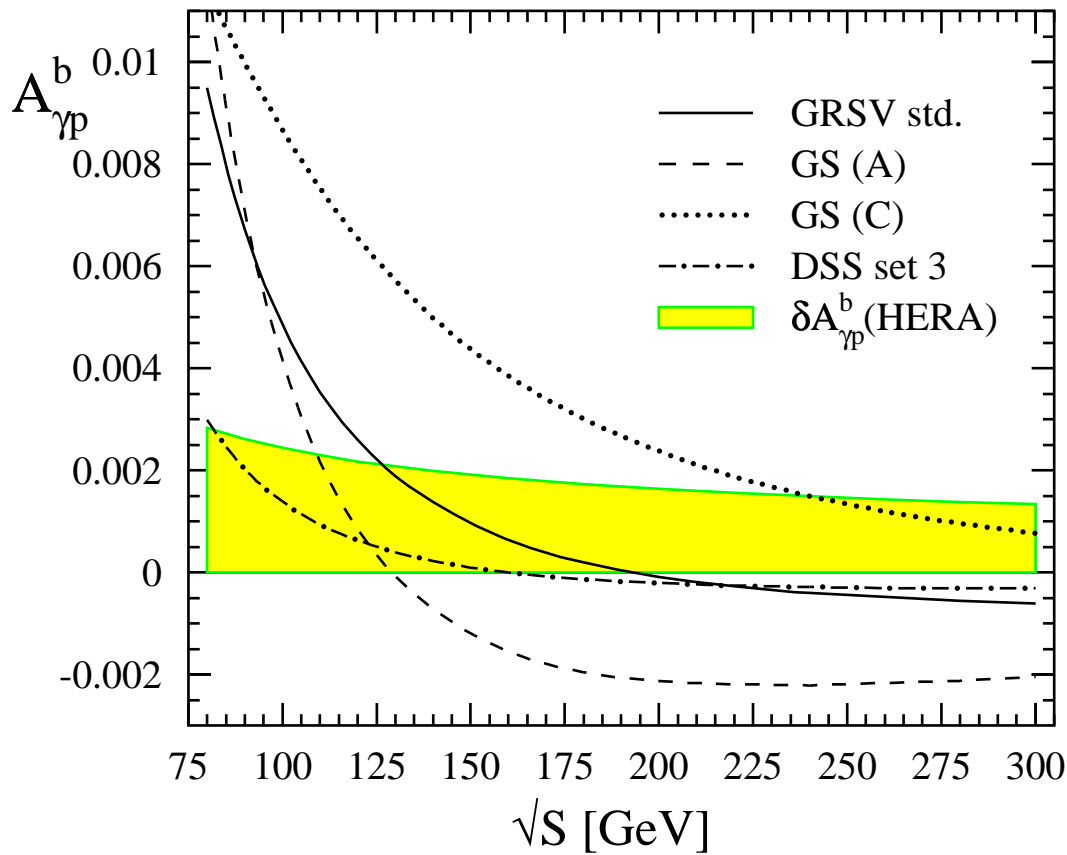}
\end{center}
\vspace{-.8cm}
\caption{\sf The longitudinal spin asymmetry $A_{\gamma p}^b$
for total bottom
photoproduction in NLO in the HERA energy range for $m=4.5\,\mathrm{GeV}$,
$\mu_f=\mu_r=2m$ and four different sets of NLO polarized
parton densities \cite{ref:grsv,ref:gs,ref:dss}.
The ``error band''  is an estimate for
the expected statistical accuracy $\delta A_{\gamma p}^b$
for such a measurement at a polarized HERA (see text).}
\end{figure}
In Fig.~9 we turn to the longitudinal spin asymmetry $A_{\gamma p}^b$
for total bottom quark photoproduction $(m=4.5\,\mathrm{GeV})$
in NLO for HERA energies for four different sets of
polarized parton distributions \cite{ref:grsv,ref:gs,ref:dss}
which mainly differ in $\Delta g$.
The results obtained for the different
sets of parton densities are well separated and sensitive to 
the different $\Delta g$, but $A_{\gamma p}^b$ 
is extremely small. Since $A_{\gamma p}^c$ already appears to be 
unmeasurable at HERA, the prospects for a meaningful measurement of 
$A_{\gamma p}^b$ seem to be not very promising at the first sight, since
bottom cross sections are smaller due to the larger $b$ quark mass 
and the smaller heavy quark charge $(e_b/e_c)^2=1/4$.
However, $b$ quarks are experimentally much easier to detect, e.g.,
through their longer lifetime (secondary vertex tag), which might
compensate these shortcomings.
The shaded band in Fig.~9 illustrates the expected statistical
accuracy for such a measurement at HERA estimated via
\begin{equation}
\label{eq:error}
\delta A_{\gamma p}^b \simeq \frac{1}{P_e P_p} 
\frac{1}{\sqrt{\varepsilon_b {\cal{L}} \sigma_{\gamma p}^b}}
\end{equation}
assuming a polarization $P_{e,p}$ of the electron and proton beams
of about $70\%$, an integrated luminosity of 
${\cal{L}}=500\,\mathrm{pb}^{-1}$ \cite{ref:herapol},
and an optimal detection efficiency
of $\varepsilon_b=0.05$ \cite{ref:yorgos}.

Finally, let us turn to some results for differential distributions.
Although their experimental relevance seems to be remote, apart
from $p_T$ and $y$ acceptance cuts, a comparison of the LO and NLO
distributions is of theoretical interest to understand in which
kinematical regions the corrections are most relevant.  

%
%
\begin{figure}[t]
\begin{center}
\vspace{-0.3cm}
\epsfig{file=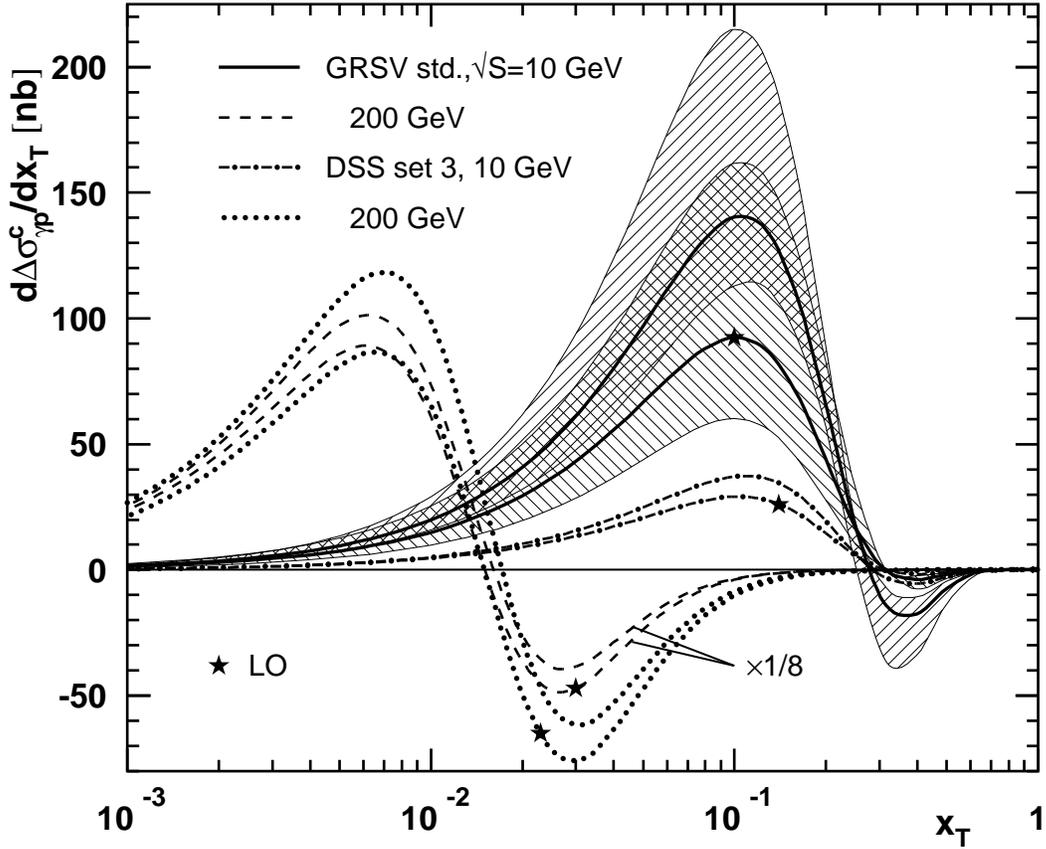}
\end{center}
\vspace{-0.2cm}
\caption{\sf $x_T$ differential polarized anticharm photoproduction cross
section $d\Delta\sigma^c_{\gamma p}/dx_T$ in LO and NLO according to
Eqs.~(\ref{eq:spics})-(\ref{eq:igrenz}) 
for $\sqrt{S}=10$ and 200 GeV and 
two sets of polarized parton distributions \cite{ref:grsv,ref:dss}. 
The bands with forward (NLO) and backward (LO) slanted hatches
correspond to the
uncertainty due to independent variations of $\mu_f$ and $\mu_r$ (see text).
LO results are denoted by stars, and
the ``GRSV std.'' curves for $\sqrt{S}=200\,\mathrm{GeV}$ are 
multiplied by $1/8$.}
\end{figure}
In Fig.~10 we show the rapidity integrated polarized cross section in
LO and NLO as a function of $x_T$ according to
Eqs.~(\ref{eq:spics})-(\ref{eq:igrenz}).  
Two values of $\sqrt{S}$ (10 and 200 GeV) and the GRSV ``standard''
\cite{ref:grsv} and DSS ``set 3'' \cite{ref:dss} polarized parton
densities are used.
For the GRSV results the theoretical uncertainty of varying
$\mu_f^2$ and $\mu_r^2$ independently in the range $a(p_T^2+m^2)$ with
$a=1/4,\ldots,4$ is illustrated by the bands with forward (NLO) and
backward (LO) slanted hatches.
All curves are calculated for the choice $a=1$ and the
LO results are marked by stars.
The NLO corrections are sizable, but the NLO shape is very similar
to the LO one. Note that the large corrections for the GRSV \cite{ref:grsv}
densities are to a large extent due to the differences between the
poorly constrained LO and NLO $\Delta g$. We have made a similar
observation for the total cross section in \cite{ref:letter};
using the GS \cite{ref:gs} densities (not shown) this effect 
would be even more pronounced. The GRSV and DSS
curves only differ in size, due to the much smaller
DSS ``set 3'' gluon, but not in shape. 
As for the total cross section in Fig.~7 it turns out that the
scale uncertainties are reduced in NLO. For example, at $x_T=0.1$ the
NLO result varies by $-20\%$ to $55\%$ whereas the LO result varies by
$-35\%$ to $75\%$.
For $\sqrt{S}=200\,\mathrm{GeV}$ the scale dependence
is similar in NLO, but even worse in LO.

%
%
\begin{figure}[t]
\begin{center}
\vspace{-0.5cm}
\epsfig{file=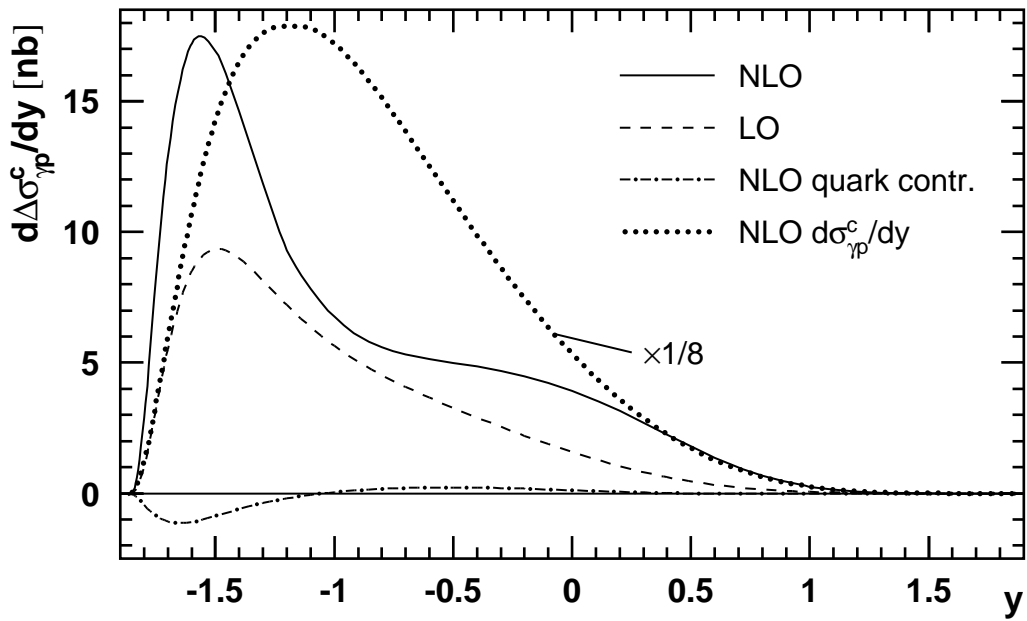}
\end{center}
\vspace{-0.6cm}
\caption{\sf \label{fig:y} Rapidity $y$ differential polarized
anticharm distribution $d\Delta\sigma^c_{\gamma p}/dy$
at $\sqrt{S}=10\,\mathrm{GeV}$ in NLO and LO
according to Eqs.~(\ref{eq:spics})-(\ref{eq:igrenz}) and
using $\mu_f=\mu_r=2m$. The light quark induced NLO contribution
is shown separately. For comparison the dotted curve displays the
NLO unpolarized distribution scaled down by a factor 8.}
\end{figure}
In Figs.~11 and 12 we show the c.m.s.\ rapidity $y$ 
and $x_F$ differential polarized anticharm photoproduction
cross sections according to Eqs.~(\ref{eq:spics})-(\ref{eq:igrenz}).
Since $p_T$ is integrated over the entire kinematical range,
we choose $\mu_f=\mu_r=2m$ as the hard scales here.
The distributions are asymmetric in $y$ and $x_F$ and the heavy quark
is dominantly produced ``backward'' with respect to the incoming
photon, i.e., in the direction of the proton.
The NLO results are always larger than the LO ones and deviate
in shape.
In both figures the unpolarized distributions,
scaled down to the size of the polarized one, are shown for comparison.
The genuine NLO contribution with light quarks in the
initial state is shown separately and appears to be negligible 
in the entire $y$ and $x_F$ range.
%
%
\begin{figure}[t]
\begin{center}
\vspace{-0.5cm}
\epsfig{file=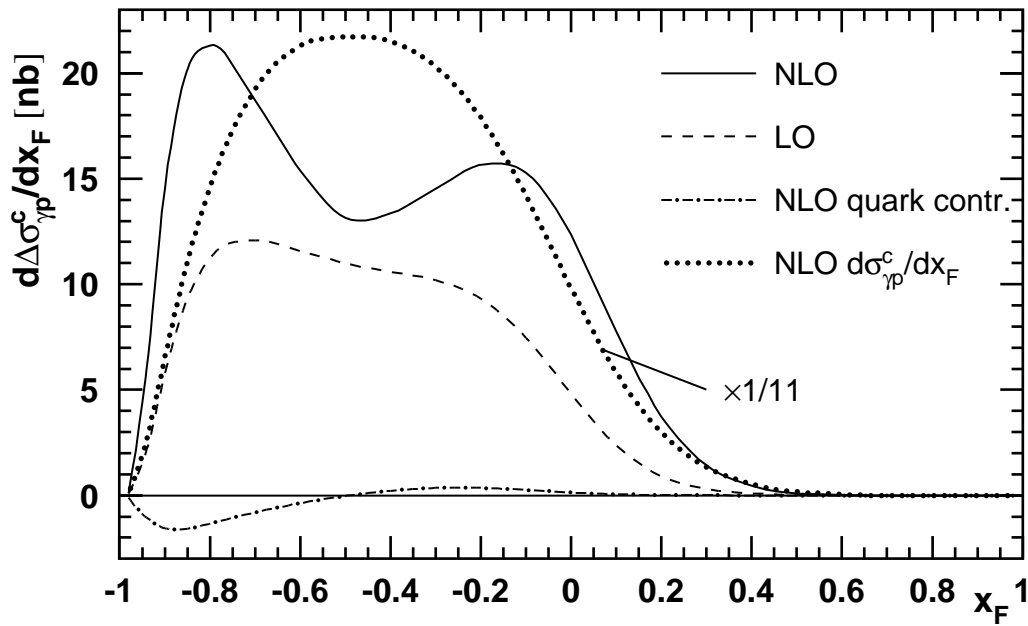}
\end{center}
\vspace{-0.5cm}
\caption{\sf \label{fig:xf} As in Fig.~11, but now as a function of $x_F$.
The NLO unpolarized result is divided by a factor $11$ here.}
\end{figure}
%

\section{Summary}
To conclude, we have presented the details of the first complete NLO
QCD calculation of heavy flavor photoproduction with longitudinally
polarized beam and target.
We have provided all relevant intermediate
steps of our calculation, in particular we have given complete
analytical results for the soft plus virtual gluon cross section. 
A compact notation was introduced to present both the unpolarized
and polarized results simultaneously, and whenever possible
we have compared our results to the existing unpolarized calculations
and found complete agreement. 
Similarly, for the abelian part of our unpolarized and polarized results, 
which can be compared to Refs.~\cite{ref:conto} analytically.

As phenomenological applications of our results we have first
explored the theoretical uncertainties due to the independent 
variation of the factorization and renormalization scales and due
the unknown precise value of the charm quark mass.
It was found that the scale dependence is much reduced in NLO, which clearly
demonstrates the usefulness of our NLO results in future 
determinations of the polarized gluon density $\Delta g$, for instance,
by the COMPASS experiment. It was critically discussed that the value
of the charm quark mass is one of the major uncertainties in a
measurement of $\Delta g$ at fixed target energies.
NLO estimates for the total bottom quark spin asymmetry accessible
in a possible future polarized collider mode of HERA were presented.
Finally, we have presented for the first time $x_T$, $y$, and
$x_F$ spin-dependent differential single anticharm photoproduction 
cross sections in NLO. Although their experimental relevance seems to be
remote, our differential expressions are useful for $p_T$ and
$y$ acceptance cuts for upcoming ``total'' cross section measurements,
in particular at COMPASS.

\section*{Acknowledgments}
I.B. is indebted to E.\ Reya for suggesting this line of inquiry 
and thanks S.\ Kretzer and I.\ Schienbein for helpful comments. 
M.S.\ is grateful to W.\ Vogelsang for valuable discussions.  
The work of I.B.\ has been supported in part by the
``Bundesministerium f\"ur Bildung, Wissenschaft, Forschung und
Technologie'', Bonn.

\newpage
\setcounter{equation}{0}
\def\theequation{A\arabic{equation}}
\section*{Appendix A: Kinematics and Hat Phase Space}
%
The phase space calculation for the $2\to 3$ processes is 
conveniently performed in the c.m.s.\ frame of the two outgoing
unobserved partons \cite{ref:gj,ref:haber,ref:smith2}
(we follow here as far as possible the notations in App.~B of
\cite{ref:smith2}):
\begin{eqnarray}
\label{eq:p1def}
\nonumber
k_3 &=& \left( \omega_3,k_3^x,\omega_3 \sin\theta_1 \cos\theta_2,
\omega_3 \cos\theta_1,\hat{k}_3\right)\;\;, \\
p_1 &=& \left(E_1,p_1^x,-\omega_3\sin\theta_1
\cos\theta_2,-\omega_3\cos\theta_1,\hat{p}_1\right)\;\;.
\end{eqnarray}
Since $k_3+p_1=0$, only one $(n-4)$-dimensional vector
$\hat{k}\equiv\hat{k}_3=-\hat{p}_1$ remains and similarly
$k_3^x = -p_1^x$. The $x$-components need not to be
specified, since the matrix elements do not depend on them, and hence
the $x$-integrations can be trivially performed.
The other three observed momenta $k_1$, $k_2$, and $p_2$ can be oriented
in such a way that they lie in the $y-z$ plane with vanishing 
hat components. There are three sets depending on which vector is chosen
to point in the $z$ direction. As outlined in Sec.~4, the whole
calculation can be performed by using only one of these sets.
``Set~I'' is given by
\begin{eqnarray}
\label{eq:set1}
\nonumber
k_1&=&\left(\omega_1,0,|\vec{p}| \sin \Psi, |\vec{p}|\cos\Psi - \omega_2,
\hat{0}\right)\;\;,\\
k_2&=&\left(\omega_2,0,0,\omega_2,\hat{0}\right)\;\;,\\
\nonumber
p_2&=&\left(E_2,0,|\vec{p}| \sin \Psi, |\vec{p}|\cos\Psi, 
\hat{0}\right)\;\;,
\end{eqnarray}
where the kinematical quantities in (\ref{eq:p1def}) and (\ref{eq:set1})
are determined from on-mass-shell constraints and
momentum conservation \cite{ref:smith2}:
\begin{eqnarray}
\label{eq:kinematics}
\nonumber
\omega_1=\frac{s+u_1}{2\sqrt{s_4+m^2}}\;\;,\;\;
\omega_2 &=& \frac{s+t_1}{2\sqrt{s_4+m^2}}\;\;,\;\;
\omega_3=\frac{s_4}{2\sqrt{s_4+m^2}}\;\;,\\
\hspace*{-2.5cm}
E_1=\frac{s_4+2m^2}{2\sqrt{s_4+m^2}}\;\;,&&\;\;
E_2=-\frac{t_1+u_1+2m^2}{2\sqrt{s_4+m^2}}\;\;,\\
\nonumber
|\vec{p}|=\frac{\sqrt{(t_1+u_1)^2-4m^2 s}}{2\sqrt{s_4+m^2}}\;\;,&&\;\;
\cos\Psi = \frac{t_1 s_4 -s (u_1+2m^2)}{(s+t_1)\sqrt{(t_1+u_1)^2-4 m^2 s}}\;\;.
\end{eqnarray}
In a second parametrization of the momenta (``set II'')
$u'$ and $u_7$ in (\ref{eq:newmandel}) are of
the [ab]-type instead of $t'$ and $u_6$ in
set I. The combined use of these two sets, depending on which type
of Mandelstam variables in the denominator has to be integrated, is
sufficient to 
avoid any appearance of auxiliary quantities like $u_6'$ introduced in 
(\ref{eq:redu}).
In set II we have
\begin{eqnarray}
\label{eq:set2}
\nonumber
k_1&=&\left(\omega_1,0,0,\omega_1,\hat{0}\right)\;\;,\\
k_2&=&\left(\omega_2,0,|\vec{p}| \sin \Psi, |\vec{p}|\cos\Psi - \omega_1,
\hat{0}\right)\;\;,\\
\nonumber
p_2&=&\left(E_2,0,|\vec{p}| \sin \Psi, |\vec{p}|\cos\Psi,
\hat{0}\right)\;\;,
\end{eqnarray}
with the same relations as in (\ref{eq:kinematics}) except for
\begin{equation}
\cos\Psi = \frac{u_1 s_4 - s (t_1 +2 m^2)}{(s+u_1)\sqrt{(t_1+u_1)^2-4m^2 s}}
\;\;.
\end{equation}

The derivation of the two-to-three body phase space formula
${\mathrm{dPS}_3}$ is 
standard, and we concentrate here only on the new aspects due to the
the additional $(n-4)$-dimensional hat-space integration in the
polarized case. The calculation of ${\mathrm{dPS}_3}$ is
facilitated by introducing
a pseudoparticle with momentum $p=p_1+k_3$, i.e., the sum of the
momenta of the two
unobserved partons. ${\mathrm{dPS}_3}$ can then be separated into a
$2\to 2$ and a $1\to 2$ phase space. Only the latter ``decay'' of the
pseudoparticle into unresolved partons depends on the
hat-space with the choice of coordinates explained above.
This non-trivial integration is then given by
\begin{equation}
\label{eq:hatps}
1\rightarrow 2 = \int d^n k_3 \, d^n p_1 \delta^+(k_3^2)
\delta^+(p_1^2-m^2) \delta^{(n)}(p-k_3-p_1)\;\;.
\end{equation}
Using (\ref{eq:p1def}) and (\ref{eq:kinematics}), i.e.,
$p=(\sqrt{s_4+m^2},0,0,0,\hat{0})$, and the fact that the matrix
elements depend only on $\hat{k}^2$,
we can evaluate (\ref{eq:hatps}) easily
by integrating $k_3$, $p_1^x$ and the angles of $\hat{k}$
\begin{eqnarray}
\label{eq:hatres}
1\rightarrow 2&=& \frac{\pi^{n/2-2}}{\Gamma (n/2-2)}
\frac{s_4^2}{8 (s_4+m^2)^{3/2}} \int_0^\pi\int_0^\pi d\theta_1 d\theta_2
\sin^2\theta_1 \sin\theta_2 \nonumber\\
&&\cdot \int_0^{\hat{k}_{max}^2}
{d}\hat{k}^2(\hat{k}^2)^{n/2-3} \left[{\frac{s_4^2}{4 (s_4+m^2)}
\sin^2\theta_1\sin^2\theta_2-\hat{k}^2}\right]^{-\frac{1}{2}}\nonumber\\
&=&\frac{\pi^{n/2-2}}{\Gamma (n/2-2)} \frac{s_4^{n-3}}{[4(s_4+m^2)]^{n/2-1}}
\int_0^\pi\int_0^{\pi} d\theta_1 d\theta_2 \sin^{n-3}\theta_1
\sin^{n-4}\theta_2\int_0^1dx \frac{x^{n/2-3}}{\sqrt{1-x}}\nonumber\\
&=&\frac{\pi^{n/2-2}}{4}\frac{\Gamma (n/2-1)}{\Gamma (n-3)}
\frac{s_4^{n-3}}{(s_4+m^2)^{n/2-1}}
\int d\Omega_{n-4}\;{\cal{I}}\;\;,
\end{eqnarray}
with the following abbreviations for the remaining integrations
\begin{eqnarray}
\label{eq:omega}
\int d\Omega_{n-4}&\equiv&
\int_0^\pi\int_0^{\pi} d\theta_1 d\theta_2  \sin^{n-3}\theta_1
\sin^{n-4}\theta_2\;\;,\\
\label{eq:idef}
{\cal{I}}&\equiv&\frac{1}{B(1/2,n/2-2)} \int_0^1dx 
\frac{x^{n/2-3}}{\sqrt{1-x}}\;\;.
\end{eqnarray}
Furthermore we have used the definition
\begin{equation}
\label{eq:xdef}
x \equiv \hat{k}^2/\hat{k}_{max}^2
=\frac{4 (s_4+m^2)\hat{k}^2}{s_4^2\sin^2\theta_1\sin^2\theta_2}\;\;.
\end{equation}

Comparing (\ref{eq:hatres}) with (B10) in Ref.~\cite{ref:smith2}, we see that 
the dependence on the hat momenta has been completely absorbed into the
additional integral ${\cal{I}}$. Thus we can schematically write
for the full $2\to 3$ phase space formula
\begin{equation}
\label{eq:dps3}
{\mathrm dPS}_3(\theta_1,\theta_2) = {\mathrm dPS}_{3,{\mathrm unp}}
(\theta_1,\theta_2) \times {\cal{I}}\;\;,
\end{equation}
where (one should keep in mind that (B14) in \cite{ref:smith2} already
includes the flux factor, etc.)
\begin{equation}
\label{eq:dps3unp}
{\mathrm dPS}_3(\theta_1,\theta_2) = dt_1 du_1\,\frac{1}{s}
\frac{(4\pi)^{-n}}{\Gamma(n-3)}
\left(\frac{t_1 u_1-m^2 s}{s}\right)^{\frac{n}{2}-1} 
\frac{s_4^{n-3}}{(s_4+m^2)^{n/2-1}}
\int d\Omega_{n-4}\;\;.
\end{equation}
Notice that (\ref{eq:hatres})-(\ref{eq:dps3unp}) agree in
the limit $m^2\to 0$ with the result presented in \cite{ref:lionel}.

The vast majority of terms in the matrix elements 
do not depend on $\hat{k}^2$ and the
rest is proportional to $\hat{k}^2$, so that one finds only two cases
($n=4+\varepsilon$): 
\begin{eqnarray}
\label{eq:psor}
{\cal{I}}\cdot 1 &\Rightarrow& {\mathrm dPS}_3(\theta_1,\theta_2) =
{\mathrm dPS}_{3,{\mathrm unp}}(\theta_1,\theta_2)\;\;,\nonumber\\
{\cal{I}}\cdot \hat{k}^2 &\Rightarrow& 
{\mathrm dPS}_3(\theta_1,\theta_2)= {\mathrm dPS}_{3,{\mathrm unp}}
(\theta_1,\theta_2) \; \varepsilon\;
\frac{s_4^2\sin^2\theta_1\sin^2\theta_2}{4(s_4+m^2)}\;\;.
\end{eqnarray}
The IR poles in our calculation stem from terms diverging as $1/s_4^2$
for $s_4\rightarrow 0$. Such poles are canceled by the factor $s_4^2$ 
in (\ref{eq:psor}). The M poles in the matrix elements have the
collinear structure $\sin^{1+\varepsilon}\theta_1/(1-\cos\theta_1)$,
and again we find that due to the factor $\sin^2\theta_1$ in
(\ref{eq:psor}) one gets finite results. Thus all hat integrations turn
out to be infrared and collinear safe. Due to the extra $\varepsilon$ in
(\ref{eq:psor}) all hat contributions are then of 
${\cal O}(\varepsilon)$ and drop out when the $\varepsilon
\rightarrow 0$ limit is taken in the end. Note that the heavy quark
mass plays the role of an infrared regulator here. Collecting the
$s_4\/$-factors in (\ref{eq:psor}) and (\ref{eq:bremsxsec}) one gets
$s_4^{3+\varepsilon}/(s4+m^2)^{2+\varepsilon /2}$. For
$m\rightarrow 0$ one then has $s_4^{1+\varepsilon /2}$, which is not
sufficient to cancel the $1/s_4^2$ contributions. Thus in the
massless case one picks up extra finite contributions 
from the hat-space integration with infrared poles \cite{ref:lionel},
whereas collinear safety still holds.

\setcounter{equation}{0}
\def\theequation{B\arabic{equation}}
\section*{Appendix B: Tensor Integrals - Some Remarks}
%
When performing the Passarino-Veltman decomposition
\cite{ref:pass} various scalar integrals appear 
which at a first glance are not listed in \cite{ref:smith2}. 
However, they can always be cast in the form of \cite{ref:smith2}.
For example, shifting the loop momentum $q\rightarrow q+p_2$
immediately yields
\begin{equation}
\label{eq:crel}
C_0(p_1,-k_1-k_2,0,m,m)=C_0(p_2,p_1,m,0,m)\;\;.
\end{equation}
In less simple cases one can easily find the necessary relations by
inspecting the Feynman parametrization of the integral. For the three
point functions $C_0(q_1,q_2,m_1,m_2,m_3)$ the Feynman parameter integral
has a denominator $[q^2-{\cal C}]^3$ with 
${\cal C}=-a b q_1^2-a c(q_1+q_2)^2-b c q_2^2+a m_1^2+b m_2^2+c m_3^2$,
where $a,~b,~c$ are functions of the two Feynman parameters with
$a+b+c=1$. We can then prove, for instance,
\begin{equation}
\label{eq:crel2}
C_0(p_1-k_1,-k_2,0,m,m)=C_0(p_1,-k_1,0,m,m)\;\;.
\end{equation}
by simply inserting the momenta and masses in ${\cal C}$
and interchanging $b\leftrightarrow c$. 
It should be noted that exploiting the freedom to re-assign the
parameters is also essential for explicitly calculating the set of 
basic scalar integrals.

When decomposing the tensorial four point functions for the box graphs
in Figs.~3 (a) and (b), one has to keep the rather lengthy intermediate
expressions as short as possible.
For the calculation of the QED-like box \cite{ref:conto}, Fig.~3 (a), 
one can show that eight of the twenty-three scalar coefficients $D_{ij}$ 
are not independent:
\begin{eqnarray}
\label{eq:kelim}
&&D_{11} = D_{12}+D_{13}\;\;,\qquad
D_{24} = (D_{21}+D_{22}-D_{23})/2\;\;,\nonumber \\
&&D_{25} = (D_{21}-D_{22}+D_{23})/2\;\;,\qquad
D_{31} = -2 D_{32}-2 D_{33}+3 D_{36}+3 D_{37}\;\;,\nonumber \\
&&D_{34} = -D_{32}-D_{33}+2 D_{36}+D_{37}\;\;,\qquad
D_{35} = -D_{32}-D_{33}+D_{36}+2 D_{37}\;\;,\\
&&D_{310} = (-D_{32}-D_{33}+D_{36}+D_{37}+D_{38}+D_{39})/2\;\;,\qquad
D_{311} = D_{312}+D_{313}\;\;.\nonumber
\end{eqnarray}
To prove this, one starts by replacing $q\to -q$ in (\ref{eq:fourp}) and
with $-q_4=q_1+q_2+q_3$ finds 
\begin{equation}
\label{eq:scald}
D_0(q_1,q_2,q_3,m_1,m_2,m_3,m_4)=D_0(q_4,q_3,q_2,m_1,m_4,m_3,m_2)\;\;.
\end{equation}
We will abbreviate this result as $D_0=\bar{D}_0$.  
Each additional power of the loop momentum in the numerator 
introduces an extra factor $(-1)$, so that 
\begin{equation}
\label{eq:minusd}
D^\alpha=-\bar{D}^\alpha\;\;,\qquad
D^{\alpha\beta}=\bar{D}^{\alpha\beta}\;\;,\qquad
D^{\alpha\beta\gamma}=-\bar{D}^{\alpha\beta\gamma}.
\end{equation}
By writing down for example the vector decomposition and comparing the 
coefficients, one obtains
\begin{equation}
\label{eq:dvec}
D_{11}=\bar{D}_{11}\;\;,\qquad
D_{12}=\bar{D}_{11}-\bar{D}_{13}\;\;,\qquad
D_{13}=\bar{D}_{11}-\bar{D}_{12}\;\;.
\end{equation}

In the case of the QED-like box one encounters 
$D_{ij}=D_{ij}(p_1,-k_1,-k_2,0,m,m,m)$ and
$\bar{D}_{ij}=D_{ij}(p_2,-k_2,-k_1,0,m,m,m)$. But the QED-like box
is symmetric with respect to $k_1\leftrightarrow k_2$ and 
$p_1\leftrightarrow p_2$. Thus $D_{ij}=\bar{D}_{ij}$ and one obtains 
the first relation of (\ref{eq:kelim}) from (\ref{eq:dvec}) and,
analogously, the other relations are derived.
For the non-abelian
box in Fig.~3 (b) on the other hand,
one has $D_{ij}=D_{ij}(-k_1,p_1,-k_2,0,0,m,m)$
and $\bar{D}_{ij}=D_{ij}(p_2,-k_2,p_1,0,m,m,0)$ and furthermore the
$k_1\leftrightarrow k_2$ symmetry is lost as discussed at the end of
Sec.~4. Thus here one finds no simple relations between the $D_{ij}$
and the $\bar{D}_{ij}$ that could be exploited to reduce the number of
coefficients. One can, of course, check these general arguments
explicitly, e.g., for arbitrary momenta and masses one finds
$D_{12}-\bar{D}_{12}=D_{13}-\bar{D}_{13}$, see (\ref{eq:dvec}).
But this difference only vanishes in the QED case,
whereas in the non-abelian case a rest
remains that is partly $t_1\leftrightarrow u_1$ antisymmetric and has
$1/\varepsilon^2$ and $1/\varepsilon$ poles. 

Finally, we note that the gluon self-energy loops
in Fig.~\ref{fig:virt} (i) are zero for massless particles (gluons and 
light quarks) due to Eq.~(\ref{eq:zero}).
For the massive quark loop one has to calculate two-point functions of 
the type $B(k_2,m,m)$ with $k_2^2=0$ and $m\neq 0$. Here problems
occur when naively applying the decomposition procedure, since
one would divide by $k_2^2$ for the coefficients $B_1$ and 
$B_{21}$. This is the simplest example of the general problem that the
standard decomposition breaks down whenever projective momenta do not
exist \cite{ref:pass}.
Of course, the self-energy integral is simple enough to be 
calculated directly and is given by
\begin{eqnarray}
\label{eq:self}
\!\!\!\!\!\!\!\Pi^{\mu\rho}(k_2^2=0,m\neq 0)
\!&=& \!i g_s^2 \frac{1}{2}\delta_{ab} \mu^{-\varepsilon}
\int\frac{d^nq}{(2\pi)^n} \frac{{\mathrm Tr}\left[\gamma^\mu (\not\!
q+m)\gamma^\rho(\not\! q+\not\! k_2+m)\right]}{(q^2-m^2)[(q+k_2)^2-m^2]}
\nonumber\\
\! &=& \! -\frac{g_s^2}{4\pi}\delta_{ab}(k_2^\mu k_2^\rho-k_2^2 g^{\mu\rho})
\frac{1}{6}\left[\frac{2}{\varepsilon}+\gamma_E-\ln (4\pi)
-\ln\left(\frac{\mu^2}{m^2}\right)\right]\;\;.
\end{eqnarray}

\setcounter{equation}{0}
\def\theequation{C\arabic{equation}}
\section*{Appendix C: Virtual plus Soft Coefficients}
%
We list here the polarized coefficients for the virtual plus soft
gluon cross section as defined in
Eq.~(\ref{eq:vssplit}):
\begin{eqnarray}
\label{eq:lqed}
\lefteqn{\Delta L_{QED}=
  \left[-t_1 (2 t_1 + u_1)/(t u_1) - u_1 (t_1 + 2 u_1)/(t_1 u)\right]/4 +
  \left[-4 m^2 s (2 t_1^2 - t_1 u_1 + 2 u_1^2) \right.} \nonumber\\
&&{}+ \left.
  t_1 u_1 (5 t_1^2 + 2 t_1 u_1 + 5 u_1^2)\right]/(4 t_1^2 u_1^2) + 
  \left\{\beta (2 m^2 s + t_1^2 + 4 t_1 u_1 + u_1^2)/(4 t_1 u_1) \right. 
  \nonumber\\ 
&&{}+ \left.
  \left[t_1^2 u_1^2 (3 t_1^2 + 4 t_1 u_1 + 3 u_1^2) + 
  4 m^4 (t_1^4 + 2 t_1^3 u_1 - 8 t_1^2 u_1^2 + 2 t_1 u_1^3 + 
  u_1^4)\right]/(4 t_1^3 u_1^3)\right\} \zeta (2) \nonumber\\
&&{}- 
  \left\{\beta (2 m^2 s + t_1^2 + 4 t_1 u_1 + u_1^2)/8 + 
  (-24 m^4 + 3 t_1^2 + 4 t_1 u_1 + 3 u_1^2)/8\right\}/ (t_1 u_1)\ln^2 \varkappa
  \nonumber\\
&&{}+
  \bigg\{-\left[2 m^2 t_1^2 (6 m^4 + 9 m^2 t_1 + 4 t_1^2) + 
  t_1 (2 m^2 + t_1) (8 m^4 + 9 m^2 t_1 + 2 t_1^2) u_1 \right. 
  \nonumber\\
&&{}+ \left.
  3 t (2 m^2 + t_1)^2 u_1^2\right]/(4 t^2 t_1^2 u_1) - 
  \left[s t_1^2 u_1 (t_1 + 2 u_1) + 
  4 m^4 (t_1^3 + 3 t_1^2 u_1 - t_1 u_1^2 - u_1^3)\right. 
  \nonumber\\
&&{}+ \left.
  2 m^2 (t_1^4 + t_1^3 u_1 - 3 t_1^2 u_1^2 - 2 t_1 u_1^3 - 
  u_1^4)\right] /(2 \beta s t_1^2 u_1^2) \ln \varkappa \bigg\}
  \ln\left(\frac{-t_1}{m^2}\right) \nonumber\\
&&{}+ 
  \bigg\{ - \left[ t_1 u_1^3 (3 t_1 + 2 u_1) + 
  4 m^6 (3 t_1^2 + 4 t_1 u_1 + 3 u_1^2) + 
  m^2 u_1^2 (15 t_1^2 + 13 t_1 u_1 + 8 u_1^2) \right. \nonumber\\
&&{}+ \left.
  2 m^4 u_1 (12 t_1^2 + 13 t_1 u_1 + 9 u_1^2)\right]/
  (4 t_1 u^2 u_1^2) - \left[ s t_1 u_1^2 (2 t_1 + u_1) - 
  4 m^4 (t_1^3 + t_1^2 u_1 \right. \nonumber\\ 
&&{}- \left.
  3 t_1 u_1^2 - u_1^3) - 
  2 m^2 (t_1^4 + 2 t_1^3 u_1 + 3 t_1^2 u_1^2 - t_1 u_1^3 - 
  u_1^4)\right] /(2 \beta s t_1^2 u_1^2) \ln \varkappa \bigg\} 
  \ln\left(\frac{-u_1}{m^2}\right) \nonumber\\ 
&&{}+ \left\{ (m^2 s - t_1 u_1)/(\beta t_1 u_1) + 
   \beta (2 m^2 s + t_1^2 + 4 t_1 u_1 + u_1^2) / (2 t_1 u_1)
   \ln(1 + \varkappa) \right\} \ln \varkappa \nonumber\\
&&{}+ 
   \left\{2 m^2 t_1^2 (u_1 - t_1) + 
   t_1^2 (2 t_1^2 + 2 t_1 u_1 + u_1^2) - 
   2 m^4 (5 t_1^2 + 2 t_1 u_1 + u_1^2) \right\} / (2 t_1^3 u_1)
   \nonumber\\
&&{}\times
   {\mathrm Li}_2 \left( \frac{t}{m^2} \right)+
   \left\{2 m^2 (t_1 - u_1) u_1^2 + 
   u_1^2 (t_1^2 + 2 t_1 u_1 + 2 u_1^2) - 
   2 m^4 (t_1^2 + 2 t_1 u_1 + 5 u_1^2) \right\}
   \nonumber\\
&&{}
   /(2 t_1 u_1^3)
   {\mathrm Li}_2 \left( \frac{u}{m^2} \right) + 
   \beta (2 m^2 s + t_1^2 + 4 t_1 u_1 + u_1^2) /(2 t_1 u_1)
   {\mathrm Li}_2(-\varkappa) + \Delta B_{QED} 
   \nonumber\\
&&{} \times
   \Bigg\{1 + \frac{s-2 m^2}{\beta s}
   \Bigg(2 \zeta (2)+ 
   \Bigg[-1 + \ln \left(\frac{-t_1}{m^2}\right) + 
   \ln\left(\frac{-u_1}{m^2}\right)+ 4 \ln(1 - \varkappa) \nonumber\\ 
&&{}-
   \ln \varkappa \Bigg] \ln \varkappa + 
   4 {\mathrm Li}_2( \varkappa)\Bigg)\Bigg\}/2\;\;,\\
\label{eq:lqedd}
\lefteqn{\Delta L_{QED}^\Delta=
  -\Delta B_{QED} 
  \left\{1 + \frac{s-2 m^2}{\beta s} \ln \varkappa\right\}
  \ln\left(\frac{\Delta}{m^2}\right)\;\;,}\\
\label{eq:lok}
\lefteqn{\Delta L_{OK}=
  m^2 s (t_1^2 + u_1^2)/(2 t_1^2 u_1^2) + 
  \bigg\{-\beta (2 m^2 s + s^2 + 2 t_1 u_1)/(4 t_1 u_1)+ 
  \bigg[-2 t_1^2 u_1^2 (2 t_1^2  }\nonumber\\
&&{}+
  t_1 u_1 + 2 u_1^2) +  m^2 s t_1 u_1 (7 t_1^2 - 8 t_1 u_1 + 7 u_1^2) - 
  m^4 (t_1^4 + 2 t_1^3 u_1 - 26 t_1^2 u_1^2 + 2 t_1 u_1^3 
  \nonumber\\ 
&&{}+  
  u_1^4)\bigg]/(2 t_1^3 u_1^3)\bigg\} \zeta (2) + 
  (-2 m^2 s + t_1^2 + u_1^2)/(4 t_1 u_1)  
  \left[ \ln^2 \left(\frac{-t_1}{m^2}\right) + 
         \ln^2 \left(\frac{-u_1}{m^2}\right) \right]
  \nonumber\\ 
&&{}-  
  \left\{ 24 m^4 - 3 s^2 + 2 t_1 u_1 - 
  \beta (2 m^2 s + s^2 + 2 t_1 u_1)\right\}/(8 t_1 u_1) 
  \ln^2 \varkappa \nonumber\\
&&{}+
  \Bigg\{(m^2 s + t_1^2) (-m^2 s + t_1 u_1)/
  (2 t t_1^2 u_1) + \left[t_1 u_1 (t_1^2 + u_1^2) - 
  2 m^2 s (2 t_1^2 - t_1 u_1 +2 u_1^2)\right]
  \nonumber\\
&&{}/
  (2 t_1^2 u_1^2) \ln\left(\frac{-u_1}{m^2}\right)
   - \left\{ s t_1^2 (s - u_1) u_1 + 
  2 m^2 \left[s t_1^3 + (s^2 + 2 t_1^2) u_1^2\right] - 
  4 m^4 \left[s^3+2 t_1 (s^2 \right. \right. 
  \nonumber\\
&&{}+ \left. \left.
   t_1 u_1)\right]\right\}/ (2 \beta s t_1^2 u_1^2)\ln \varkappa
   \Bigg\}\ln\left(\frac{-t_1}{m^2}\right)  + 
   \Bigg\{(-m^2 s + t_1 u_1) (m^2 s + u_1^2)/ (2 t_1 u u_1^2)
   \nonumber\\
&&{}-
  \left\{s (s - t_1) t_1 u_1^2 - 
  4 m^4 \left[s^3 + 2 u_1 (s^2 + t_1 u_1)\right] + 
  2 m^2 \left[s u_1^3 + t_1^2 (s^2 + 2 u_1^2)\right]\right\} 
  /(2 \beta s t_1^2 u_1^2)
   \nonumber\\
&&{}\times
   \ln \varkappa \Bigg\} \ln\left(\frac{-u_1}{m^2}\right) + 
   \left\{-(m^2 s - t_1 u_1)/(\beta t_1 u_1) - 
   \beta (2 m^2 s + s^2 + 2 t_1 u_1)/(2 t_1 u_1)
   \right. \nonumber\\
&&{}\times \left. 
  \ln(1 + \varkappa)\right\} \ln \varkappa + 
  \left\{m^2 t_1^2 (t_1 - 3 u_1) - t_1^3 (t_1 + 2 u_1) + 
  m^4 (5 t_1^2 + 2 t_1 u_1 + u_1^2)\right\} / (2 t_1^3 u_1)
  \nonumber\\
&&{}\times
 {\mathrm Li}_2\left(\frac{t}{m^2}\right) + 
  \left\{m^2 u_1^2 (-3 t_1 + u_1) - 
  u_1^3 (2 t_1 + u_1) + m^4 (t_1^2 + 2 t_1 u_1 + 5 u_1^2)\right\}
  /(2 t_1 u_1^3) \nonumber\\
&&{}\times {\mathrm Li}_2\left(\frac{u}{m^2}\right)- 
  \beta (2 m^2 s + s^2 + 2 t_1 u_1) /(2 t_1 u_1)
  {\mathrm Li}_2(-\varkappa) + \Delta B_{QED} \Bigg\{\Bigg[-3 \zeta (2)
  \nonumber\\
&&{} +
  4 \ln \left(\frac{\mu_f^2}{m^2}\right) 
  \ln\left(\frac{-u_1}{m^2}\right) + 
  \ln^2\left(\frac{t_1}{u_1}\right) - 
  2 \ln\left(\frac{t_1}{u_1}\right) \ln \varkappa - \ln^2 \varkappa 
  + 2 {\mathrm Li}_2 \left(1 - \frac{t_1}{u_1 \varkappa}\right)
  \nonumber\\
&&{}- 
  2 {\mathrm Li}_2 \left(1 - \frac{u_1}{t_1 \varkappa}\right)\Bigg]/4 - 
  \frac{s-2 m^2}{\beta s} \Bigg[2 \zeta (2) + 
  \Bigg\{ \ln\left(\frac{-t_1}{m^2}\right) + 
          \ln\left(\frac{-u_1}{m^2}\right)
  \nonumber\\
&&{}+ 
  4 \ln(1 - \varkappa) - \ln \varkappa\Bigg\} \ln \varkappa 
  + 4 {\mathrm Li}_2(\varkappa)\Bigg]/2\Bigg\}
\;\;,\\
\label{eq:lokd}
\lefteqn{\Delta L_{OK}^\Delta=
  \Delta B_{QED} \!
  \left\{\! \ln\left(\frac{\Delta}{m^2}\right) - 
  \ln\left(\frac{\mu_f^2}{m^2}\right) + 
  \ln\left(\frac{t_1}{u_1}\right) 
  + \frac{s-2 m^2}{\beta s} \ln \varkappa \right\}\!
 \ln\left(\frac{\Delta}{m^2}\right)\;\;,}\\
\label{eq:lrf}
\lefteqn{\Delta L_{RF}= \Delta B_{QED}
 \ln\left(\frac{\mu_r}{\mu_f}\right) \;\;,}
\end{eqnarray}
with $\Delta B_{QED}$ given in Eq.~(\ref{eq:lores}), 
$\beta=\sqrt{1-4m^2/s}$, $\varkappa\equiv(1-\beta)/(1+\beta)$,
and $t$ and $u$ as defined in (\ref{eq:lomandel}).

When integrated, the hard gluon cross section diverges 
logarithmically $\sim \ln^k \Delta/m^2$ $(k=1,\,2)$ as the
IR cutoff $\Delta\rightarrow 0$. 
This is by definition canceled by the sum of the
$\tilde{L}^\Delta$ contributions in the soft gluon cross section.
If one is interested to show the contributions from the hard and the
soft plus virtual contributions separately
(as, e.g., in Fig.~1 of \cite{ref:letter}), it is thus advisable
to shift the $\tilde{L}^\Delta$ terms to the, in this way
redefined, hard cross section in order to achieve a numerical
stable result independent of $\Delta$. 
For any numerical calculation of physically relevant hadronic 
cross sections, it is however more practical to directly add the 
complete soft plus virtual piece to the hard cross section.
In both cases this can be achieved by rewriting the soft plus virtual
cross section, expanded in powers of 
$\ln^k \Delta/m^2$ ($k=0,\,1,\,2)$, as follows \cite{ref:smith3}
\begin{equation}
\label{eq:delshift}
\delta(s_4) \sum_{k=0}^2\alpha_k \ln^k \frac{\Delta}{m^2}
\rightarrow \Theta(s_4-\Delta) A_k \alpha_k\left|_{s_4=0} \right.
\end{equation}
with certain coefficients $\alpha_k$. 
(\ref{eq:delshift}) takes properly care of the different distributions 
$\delta (s_4)$ and $\Theta (s_4-\Delta)$ multiplying the soft and hard 
parts, respectively, see Eq.~(\ref{eq:s4int}). As indicated in
(\ref{eq:delshift}), the
$\alpha_k$ have to be always evaluated using the ``elastic'' 
$2\to 2$ kinematics, i.e., $s_4=0$, even when added to the 
$2\to 3$ hard cross section. The coefficients $A_k$ are given by
\begin{equation}
\label{eq:dela}
A_0=\frac{1}{s_4^{\mathrm max}-\Delta},\qquad
A_1=\frac{\ln (s_4^{\mathrm max}/m^2)}{s_4^{\mathrm max}-\Delta}
-\frac{1}{s_4},\qquad
A_2=\frac{\ln^2 (s_4^{\mathrm max}/m^2)}{s_4^{\mathrm max}-\Delta}
-\frac{2\ln (s_4/m^2)}{s_4}
\end{equation}
as can be easily verified by integrating the l.h.s.\ of (\ref{eq:delshift})
over $(\int_{\Delta}^{s_4^{\mathrm max}} ds_4)$, which recovers the
$\ln^k \Delta/m^2$ terms. 
%
%

%
\end{document}